\documentclass[trackchanges,twocolumn]{aastex701}
\usepackage{amsmath}
\DeclareUnicodeCharacter{2212}{-}
\shortauthors{Haldar, Ram, Ghosh et al.}

\begin{document}

\title{Probing the QPO and spectro-temporal evolution of MAXI J1750−-327 with XPoSat during its first detected outburst : Constraints on black-hole mass and spin}

\author[orcid=0009-0009-3019-6220,sname='haldar']{Meghamani Haldar}
\altaffiliation{These authors contributed equally to this work.}
\affiliation{Department of Physics, University of North Bengal, Siliguri, 734013, India}
\email{rs\_meghamani@nbu.ac.in}  

\author[orcid=0000-0002-3823-1459,sname='ram']{Biki Ram} 
\altaffiliation{These authors contributed equally to this work.}
\affiliation{ Department of Astronomy, Astrophysics and Space Engineering (DAASE), Indian Institute of Technology (IIT) Indore, Khandwa Road, Simrol, Indore, 453552, India}
\email{bikiram436@gmail.com}

\author[sname='ghosh']{Shubhrangshu Ghosh}
\affiliation{Center for Astrophysics, Gravitation and Cosmology (CAGC), Shri Ramasamy Memorial (SRM) University Sikkim, 5th Mile Tadong, Gangtok, 737102, India}
\affiliation{Department of Physics, Shri Ramasamy Memorial (SRM) University Sikkim, 5th Mile Tadong, Gangtok, 737102, India}
\email{shubhrangshughosh.r@srmus.edu.in}

\author[orcid=0000-0002-9738-8528,sname='chaudhury']{Kishor Chaudhury}
\affiliation{Department of Physics, University of North Bengal, Siliguri, 734013, India}
\affiliation{Department of Physics, Alipurduar University, Alipurduar, 736122, India}
\email{chaudhurykishor@gmail.com}

\author[orcid=0000-0002-3915-4360,sname='dey']{Rajat K. Dey}
\affiliation{Department of Physics, University of North Bengal, Siliguri, 734013, India}
\email{rkdey2007phy@nbu.ac.in}

\author[orcid=0000-0002-4444-7299]{Tamal Sarkar}
\affiliation{High Energy $\&$ Cosmic Ray Research Centre (HECRRC), University of North Bengal, Siliguri, 734013, India}
\email{tsarkar@nbu.ac.in}

\correspondingauthor{Shubhrangshu Ghosh \\ (shubhrangshughosh.r@srmus.edu.in) \\ \medskip Biki Ram (bikiram436@gmail.com)}

\begin{abstract}

We present a comprehensive timing and broadband spectral study of the newly discovered black hole low-mass X-ray binary candidate MAXI~J1750$-$327 during its August 2026 outburst, using 11 continuous observations obtained with the \textit{XSPECT} instrument onboard \textit{XPoSat}. 
During epochs 1$-$2, we detect prominent type C QPO exhibiting a systematic upward frequency drift, accompanied by a prominent low-frequency noise component, with a clear transition from type-C to type-B, with subsequent disappearance of QPOs after epoch 4. Broadband spectral modeling
reveals that this timing transition is directly mirrored by a fundamental restructuring of the accretion geometry. The inner disk temperature abruptly increases before settling into a steady range, accompanied by a steep reduction in disk normalization across the transition from epoch 3 to 4. This behavior signifies an active dynamical inward migration of the thin accretion disk toward the innermost stable circular orbit (ISCO). 
The Comptonized flux declines sharply after the QPO turns off, suggesting a transition from a hard to soft state, through intermediate states. We also observed that the spectro-temporal correlations provide compelling physical support for the Lense$-$Thirring (LT) precession framework to be the origin of the type-C QPOs. 
Using a Bayesian framework, we constrain the black hole mass to be $M_{\rm BH}=7.92^{+6.19}_{-3.72} M_{\odot}$ and the dimensionless spin parameter to be $a_*=0.163^{+0.116}_{-0.056}$  by jointly modeling the observed type-C QPO frequencies with the global LT precession model and marginalizing over the uncertainties in distance, inclination, and diskbb normalization.

\end{abstract}

\keywords{\uat{Black hole physics}{159} --- \uat{Accretion}{14} --- \uat{X-ray astronomy}{1810} --- \uat{Low-mass x-ray binary stars }{939}}


\section{Introduction} 
Low-mass X-ray binaries (LMXBs) consist of a stellar mass black hole (BH),  accreting matter from a low-mass ($\lesssim 1~M_{\odot}$) companion star via Roche-lobe overflow, with the transferred material forming an accretion disk around the compact source \citep{ 2006ARA&A..44...49R}.
Over the course of an outburst, the source evolves through a well-established sequence of spectral states, canonically described as the hard state, the hard-intermediate state, the soft-intermediate state, and the soft state, each characterized by a distinct combination of energy spectral shape and fast X-ray variability \citep{2005ApJ...623..383H, 2016ASSL..440...61B}. In the hard state, the energy spectrum is dominated by a Comptonized power law with a photon index of $\Gamma \approx 1.5$--$1.7$ and a high-energy cutoff around $100$--$200$~keV, attributed to inverse-Compton scattering of soft disk photons in a hot, optically thin corona, while the accretion disk itself is thought to be truncated at some radius larger than the innermost stable circular orbit (ISCO), or otherwise only weakly contributing to the observed flux \citep{2007A&ARv..15....1D, 2010MNRAS.403...61D}. As the outburst progresses, the source typically softens: the disk component strengthens and its inner radius is inferred to move inward, the coronal photon index steepens toward $\Gamma \approx 2$--$2.5$, and the source ultimately enters the soft state.
This overall evolution is conventionally visualized in a hardness-intensity diagram (HID), in which the source traces out a characteristic hysteresis-like ``q''-shaped track: rising in luminosity at roughly constant hardness along the hard-state branch, softening rapidly at high luminosity as it crosses to the soft-state branch, and eventually returning to the hard state at comparatively lower luminosity as the outburst decays \citep{2004MNRAS.355.1105F, 2005ApJ...623..383H, 2016ASSL..440...61B}. 
These spectral state transitions are accompanied by systematic changes in the variability of the X-ray flux, most notably in the fractional rms of the broadband noise and the strength of quasi-periodic oscillations \citep{2011MNRAS.418.2292M,2026MNRAS.549ag970R}.
Combined spectral and timing analysis is therefore a key diagnostic of how the accretion geometry, and in particular the innermost regions of the flow, evolves during an outburst \citep{2006AdSpR..38.2675V, 2019NewAR..8501524I}.

A defining timing signature of BH LMXBs is the appearance of low-frequency quasi-periodic oscillations (QPOs) in the power density spectrum, first identified in the early RXTE era \citep{1991ApJ...383..784M, 1989ARA&A..27..517V} and since observed to be ubiquitous during outbursts. These QPOs are conventionally classified into types A, B and C based on their centroid frequency, quality factor, rms amplitude and the strength of the accompanying broadband noise \citep{1999ApJ...526L..33W, 2005ApJ...629..403C}. Type-C QPOs, the most commonly observed variety, appear predominantly in the hard and hard-intermediate states as strong, narrow peaks with frequencies ranging from $\sim 0.01$--$30$~Hz, often accompanied by one or more harmonics, superposed on strong band-limited noise; type-B QPOs are instead confined to the soft-intermediate state near the state transition and are associated with weaker underlying noise; and type-A QPOs are weak, broad features occasionally seen in the soft state \citep{2005ApJ...629..403C, 2015MNRAS.447.2059M, 2019NewAR..8501524I}.

The physical origin of these low-frequency QPOs remains debated, but the most widely favored explanation for type-C QPOs invokes geometric Lense--Thirring precession of a hot, radially extended inner accretion flow that is misaligned with the BH spin axis \citep{1998ApJ...492L..59S,2009MNRAS.397L.101I}. In this relativistic precession model, the entire inner flow precesses rigidly at the nodal precession frequency appropriate to its outer edge, naturally producing the observed correlations between QPO frequency, spectral hardness and characteristic disk/corona radii, and explaining the strong dependence of QPO amplitude on the system's inclination \citep{2009MNRAS.397L.101I, 2015MNRAS.447.2059M, 2019NewAR..8501524I}. Alternative models attribute the oscillation to intrinsic instabilities of the accretion flow itself, such as the accretion-ejection instability driven by spiral density waves at the inner edge of a magnetized disk \citep{2002A&A...387..487R}, or to non-linear feedback between the corona and the disk that drives limit-cycle-like oscillations in the Comptonized flux \citep{2010MNRAS.404..738C,2026A&A...710A.203M}.

As the source softens and the inner flow geometry evolves toward a more disk-dominated configuration, the type-C QPO is generally observed to weaken and eventually disappear\citep{2027JHEAp..5600738R}, typically in the soft-intermediate state coincident with, or shortly before, the appearance of a type-B QPO and the transition to the soft state \citep{2005ApJ...629..403C, 2011MNRAS.418.2292M}. In the precession framework, this QPO ``turn-off'' is interpreted as the shrinking of the precessing region as the geometrically thin disk moves inward, so that the characteristic precession frequency increases and the fractional rms of the oscillation drops below detectability \citep{2009MNRAS.397L.101I, 2019NewAR..8501524I}.

The hard-to-soft transition is frequently accompanied by discrete flaring activity and enhanced variability near the state boundary, along with a marked change in the source's radio properties: the steady, compact jet associated with the hard state is quenched as the source crosses into the soft state, often preceded by a transient, optically thin radio flare linked to the ejection of discrete jet knots \citep{2004MNRAS.355.1105F, 2009MNRAS.396.1370F}. Tracking the joint evolution of the QPO frequency and amplitude, the broadband X-ray spectrum, and any associated radio flaring therefore provides a comprehensive picture of how the accretion flow reconfigures itself over the course of an outburst \citep{2025ApJ...984L..53W,2026arXiv260902681M}. 
The disappearance of the type-C QPO is therefore especially significant, since it may signal an abrupt restructuring of the precessing inner flow rather than a smooth drift in frequency\citep{2027JHEAp..5600738R}. Rapid QPO transitions, including the sudden loss of the type-C QPO, have been reported in several BHXRBs and are often accompanied by changes in the broadband noise and the emergence of a type-B QPO \citep{2021ApJ...911..127S,2023MNRAS.521.3570Y}. In Swift J1727.8–1613, a brief hard-to-soft excursion saw the type-C QPO vanish together with a marked drop in broadband noise, while a type-B QPO appeared alongside discrete radio-jet ejections evidence for a close link between the breakdown of the precessing inner flow and the accretion–ejection process \citep{2025A&A...697A.120J}.
Despite this observational progress, the physical trigger for the type-C QPO's disappearance remains poorly understood. 

Along with the spectral and timing evolution constraining the mass and spin of BHs in X-ray binaries is essential for understanding accretion dynamics and the role of spin in relativistic phenomena. The mass is best determined dynamically from optical observations of the companion, whereas the spin is commonly estimated through continuum fitting and X-ray reflection spectroscopy, both of which rely on modeling the innermost accretion flow \citep{2011CQGra..28k4009M,2014SSRv..183..277R}. These approaches are difficult for transient systems with poor optical constraints or limited high quality X-ray spectra. Low frequency QPOs (LFQPOs) offer a complementary probe of the inner accretion geometry. In particular, type-C QPOs are widely associated with Lense--Thirring (LT) precession of a geometrically thick, radially extended hot inner flow around a spinning BH \citep{2009MNRAS.397L.101I,2012MNRAS.419.2369I,2019NewAR..8501524I}. In this picture, the precession frequency depends on the BH mass and spin, as well as on the radial extent and surface-density profile of the precessing flow. The evolution of the QPO frequency therefore traces changes in the accretion geometry and can constrain the properties of the compact object. Related timing-based approaches, such as the relativistic precession model applied to simultaneous QPO measurements, have also been used to estimate BH mass and spin \citep{2014MNRAS.444.2065I,2014MNRAS.437.2554M}. 

In this work, we use XPoSat observations of the newly discovered BH transient MAXI~J1750$-$327, obtained from 2026 August 20 onward, to report the evolution of its QPO and the accompanying spectral state transition, combining timing and spectral analysis to characterize the source's outburst evolution. 
We combine the spectral evolution of the thermal disk with a global LT precession model for the type-C QPOs of MAXI~J1750$-$327, using the disk normalization to set the radial extent of the precessing flow and the ISCO corresponding to a Kerr BH as its inner boundary, to constrain its mass and spin.

\subsection{The source MAXI J1750--327}
MAXI J1750--327 is a newly discovered bright X-ray transient that was first detected by the MAXI/GSC nova alert system at 02:41 UT on 2026 August 13, at R.A. = 267.736° and Dec. = 
-32.734° (J2000) \citep{2026ATel17971....1R}. The source exhibited an average 4–10 keV flux of $117\pm12$ mCrab during the MAXI scan \citep{2026ATel17963....1N}. On the same day,
the Wide-field X-ray Telescope (WXT) detected the transient and reported a hard X-ray spectrum with a photon index of $\sim1.5$ and an absorbing column density of $N_{\rm
H}\sim1.2\times10^{22}$ cm$^{-2}$ \citep{2026ATel17964....1H}. Follow-up observations with the Einstein Probe FXT on August 14 revealed a prominent QPO at $\sim0.45$ Hz, with an FWHM of 0.08 Hz and an rms amplitude of $\sim7\%$, accompanied by a first harmonic at $\sim0.94$ Hz
\citep{2026ATel17969....1H}.
Multi-wavelength observations further supported the classification of MAXI J1750--327 as a BH X-ray binary candidate; in particular, MeerKAT detected a $\sim14$ mJy radio counterpart at 1.28 GHz on August 15, with an inverted radio spectrum 
($\alpha=+0.33\pm0.06$), consistent with a compact jet associated with the hard state \citep{2026ATel17979....1C}. Meanwhile, the evolution of the MAXI hardness ratios, particularly the decline of the 10–20 keV/2–4 keV ratio from $\sim0.55$ to $\sim0.40$, indicated that the source was undergoing a hard-to-soft state transition \citep{2026ATel17974....1N}. 
All these results suggest this source to be a BH LMXB.


\begin{table*}[!htbp]
\centering
\small
\begin{tabular}{ccccccc}
\hline
\hline
 \textbf{Observation ID} & \textbf{Epoch} & \textbf{Observation Date} & \textbf{MJD} & \textbf{Start} & \textbf{End} & \textbf{Exposure (s)} \\
 & & \textbf{(YYYY-MM-DD)} && \textbf{Time} & \textbf{Time} \\
\hline
 & 1 & 2026-08-20 & 61272 & 00:26:31 & 20:02:38 & 7313 \\
 & 2 & 2026-08-21 & 61273 & 00:45:18 & 21:57:55 & 7484 \\
 & 3 & 2026-08-22 & 61274 & 01:05:34 & 19:01:32 & 6224 \\
 & 4 & 2026-08-23 & 61275 & 01:25:06 & 19:19:35 & 6283 \\
 & 5 & 2026-08-24 & 61276 & 04:33:38 & 18:00:24 & 6979 \\
X01\_XPC\_C26\_0005 & 6 & 2026-08-25 & 61277 & 00:20:08 & 23:10:03 & 7481 \\
 & 7 & 2026-08-26 & 61278 & 00:39:32 & 23:28:03 & 7277 \\
 & 8 & 2026-08-27 & 61279 & 00:59:16 & 23:46:03 & 6224 \\
 & 9 & 2026-08-28 & 61280 & 01:19:00 & 19:22:27 & 6383 \\
 & 10 & 2026-08-29 & 61281 & 04:28:08 & 17:53:15 & 5618 \\
 & 11 & 2026-08-30 & 61282 & 00:14:10 & 23:00:18 & 5905 \\
\hline
\end{tabular}
\caption{Details of XSPECT observations over eleven different epochs during the outburst of MAXI~J1750$-$327.}
\label{tab:obs_log}
\end{table*}

\begin{figure}[ht!]
    \centering
    \includegraphics[width=1.0\linewidth]{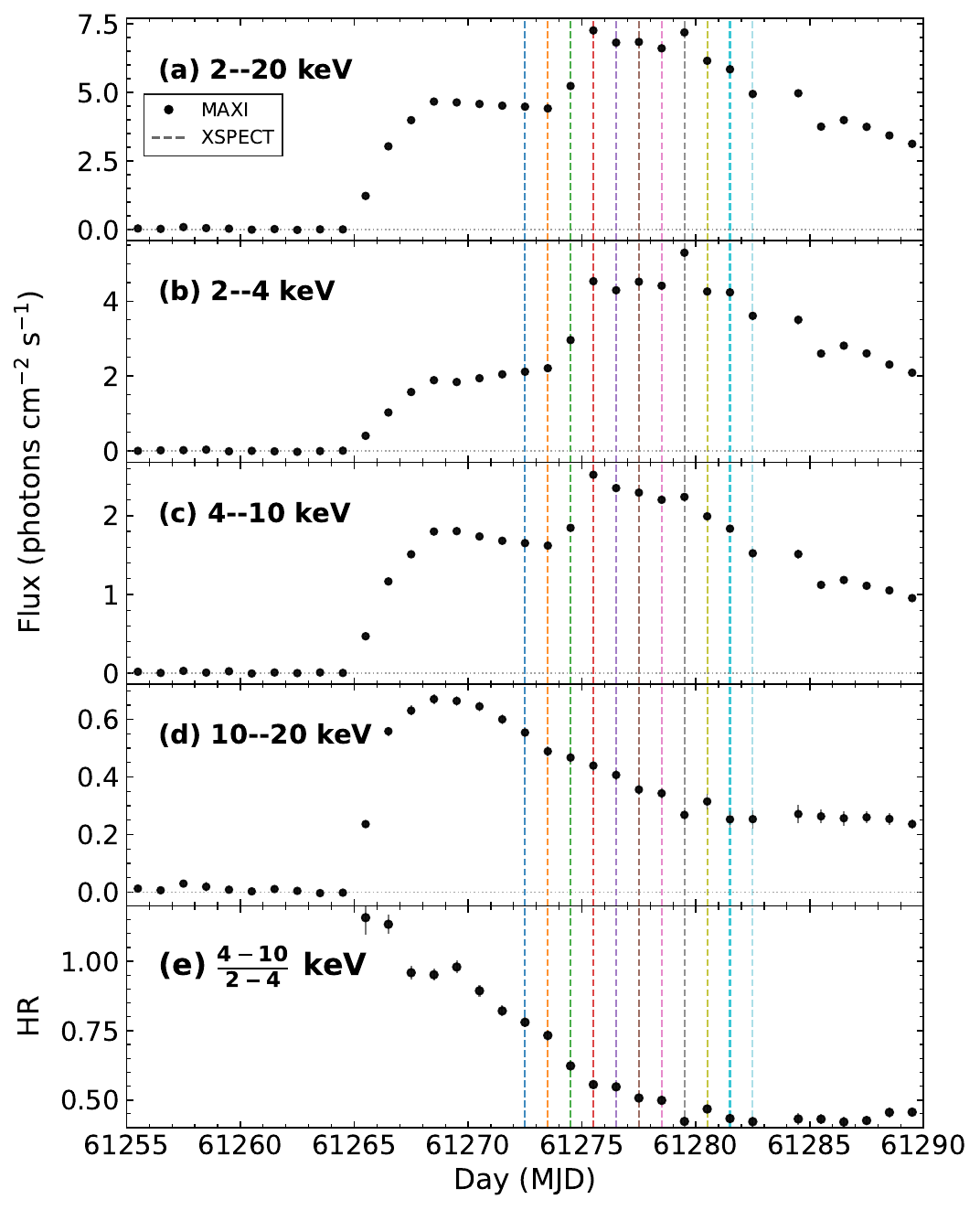}
    \caption{Representative MAXI light curves in the (a) $2$--$20\text{ keV}$, (b) $2$--$4\text{ keV}$, (c) $4$--$10\text{ keV}$, and (d) $10$--$20\text{ keV}$ ranges, along with (e) the hardness ratio (HR), are shown for MAXI~J1750$-$327. The vertical colored dashed lines mark the mid time of the 11 consecutive XSPECT observations during the August 2026 outburst. The HR is defined as the ratio between the flux in $4$--$10\text{ keV}$ and $2$--$4\text{ keV}$ energy ranges.}
    \label{Fig:maxi_lc}
\end{figure}

\section{Observation and data reduction \label{Sec:2}}
In this work, we investigate the broadband soft X-ray properties of MAXI J1750$-$327 using observations from the X-ray Spectroscopy and Timing (\emph{XSPECT}) payload onboard India's X-ray Polarimeter Satellite (\textit{XPoSat}) mission, launched in 2024. We used the first 11 epochs between 2026 August 20 and 2026 August 30 (MJD 61272$-$61282; Proposal ID: \texttt{X01\_XPC\_C26\_0005}), providing an effective total exposure time of approximately 70~ks. A complete log detailing each observation is summarized in Table~\ref{tab:obs_log}. These datasets allow us to track the spectral state evolution across all epochs, alongside the timing characteristics of the QPOs detected in the earlier phases.

Data reduction was carried out using the standard processing pipeline \texttt{xspect\_pipeline\_l1l2} (\texttt{v2.2.1}; \cite{2026MNRAS.tmp..908C}). Initial event cleaning was performed on the raw Level-1 (L1) products via \texttt{xspl2screen} to generate Level-2 (L2) event files based on the standard Good Time Interval (GTI) criteria. 
Higher-level scientific products were subsequently extracted using the \texttt{xspl2specgen} and \texttt{xspl2lcgen} modules.  The XSPECT payload consists of 16 Swept Charge Devices (SCDs) grouped under two distinct collimator fields of view (FoV): $2^\circ \times 2^\circ$ and $3^\circ \times 3^\circ$. Following \citet{2026arXiv260721365N}  we selected the $2^\circ \times 2^\circ$ FoV detectors for high-resolution timing analysis, 
of MAXI J1750$-$327 to ensure spectro-temporal consistency. 
Light curves and corresponding background products were generated using the \texttt{-backlcgen} task in the total ($0.8\text{--}11.0\,\mathrm{keV}$), soft ($0.8\text{--}4.0\,\mathrm{keV}$), and hard ($4.0\text{--}11.0\,\mathrm{keV}$) energy bands with time binnings of $50\,\mathrm{s}$ and $64\,\mathrm{s}$ for hardness-intensity and variability studies, and at a high time resolution of $0.025\,\mathrm{s}$ in the full band for power density spectral (PDS) analysis.
 Non-source background contributions were modeled using the observation-specific ``hybrid'' background scheme (\texttt{-backspecgen = True}), and ancillary response files (ARFs) accounting for our specific detector selection were simultaneously generated via \texttt{-arfgen True}.

\begin{figure}[ht!]
\includegraphics[width=\columnwidth]{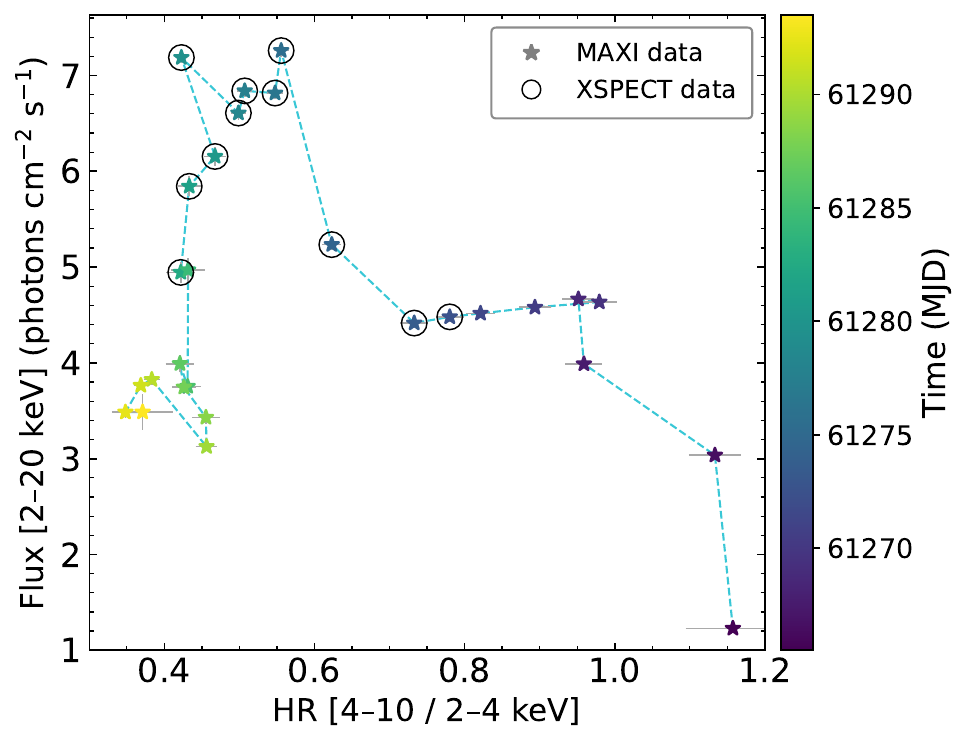}
\caption{MAXI Hardness-intensity diagram of MAXI~J1750$-$327 during the August 2026 outburst. Each point in the HID corresponds to a 1-day binned MAXI observation. The $y$-axis is taken as the flux in the $2\text{-}20\ \mathrm{keV}$ energy band, while the hardness ratio (HR) is defined as the ratio between fluxes in the $4\text{-}10\ \mathrm{keV}$ and $2\text{-}4\ \mathrm{keV}$ bands. Data points are color-coded according to the observation time (MJD). The circles denote the XSPECT observations across epochs 1–11.
\label{Fig:HID}}
\end{figure}

\begin{figure}[ht!]
\includegraphics[width=\columnwidth]{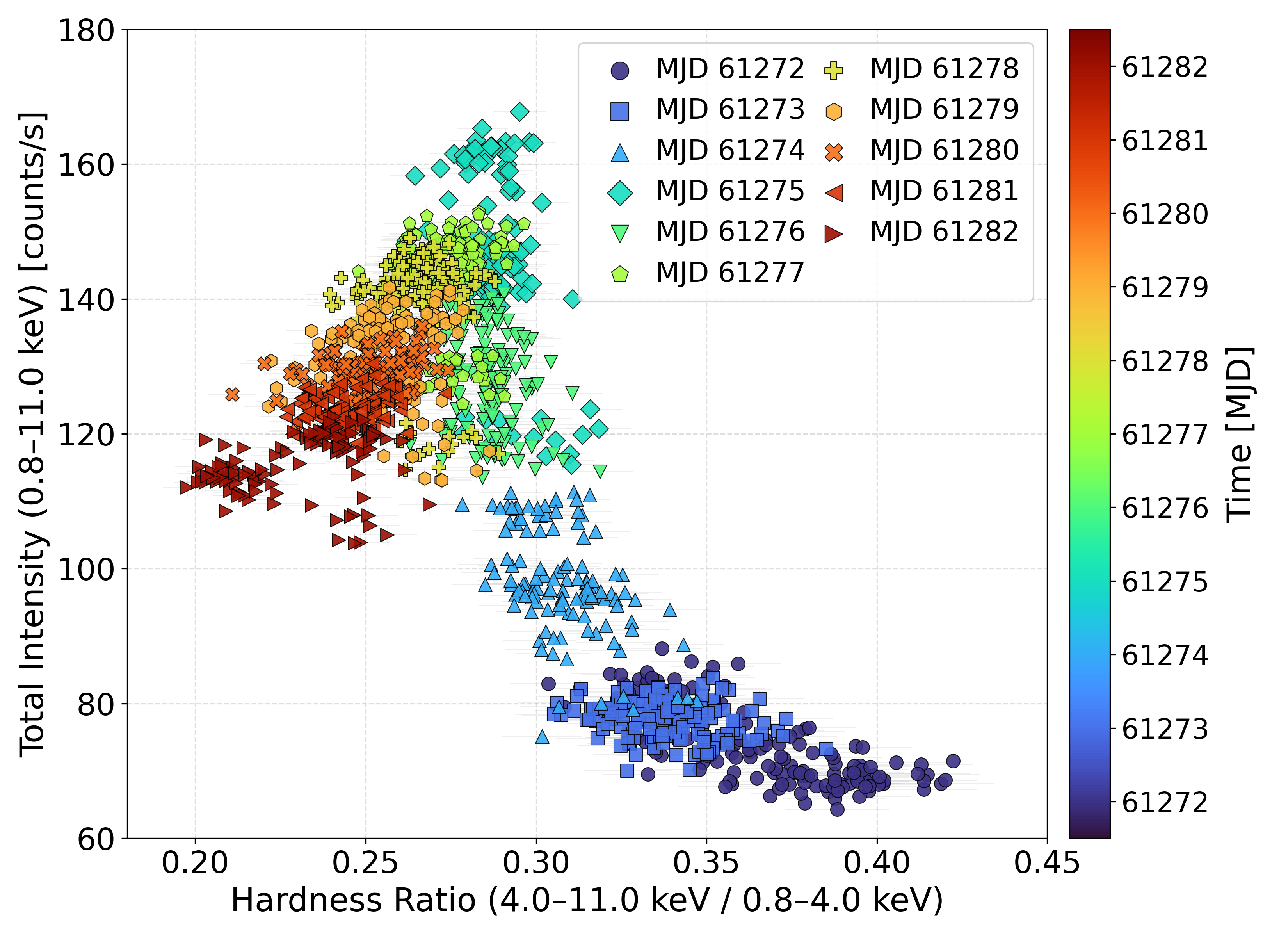}
\caption{Hardness-intensity diagram (HID) of MAXI~J1750$-$327 obtained from XSPECT observations during the 2026 August outburst (MJD 61272--61282). Each point corresponds to a $50\,\mathrm{s}$ binned light curve. The $y$-axis represents the total source intensity in the $0.8\text{--}11.0\,\mathrm{keV}$ energy band, while the hardness ratio (HR) along the $x$-axis is defined as the ratio between count rates in the $4.0\text{--}11.0\,\mathrm{keV}$ and $0.8\text{--}4.0\,\mathrm{keV}$ bands. Data points are color-coded chronologically and plotted with distinct marker shapes corresponding to each observation epoch (MJD).
\label{Fig:XSPECT_HID}}
\end{figure}

\begin{figure*}[ht!]
    \centering
    \includegraphics[width=0.8\linewidth]{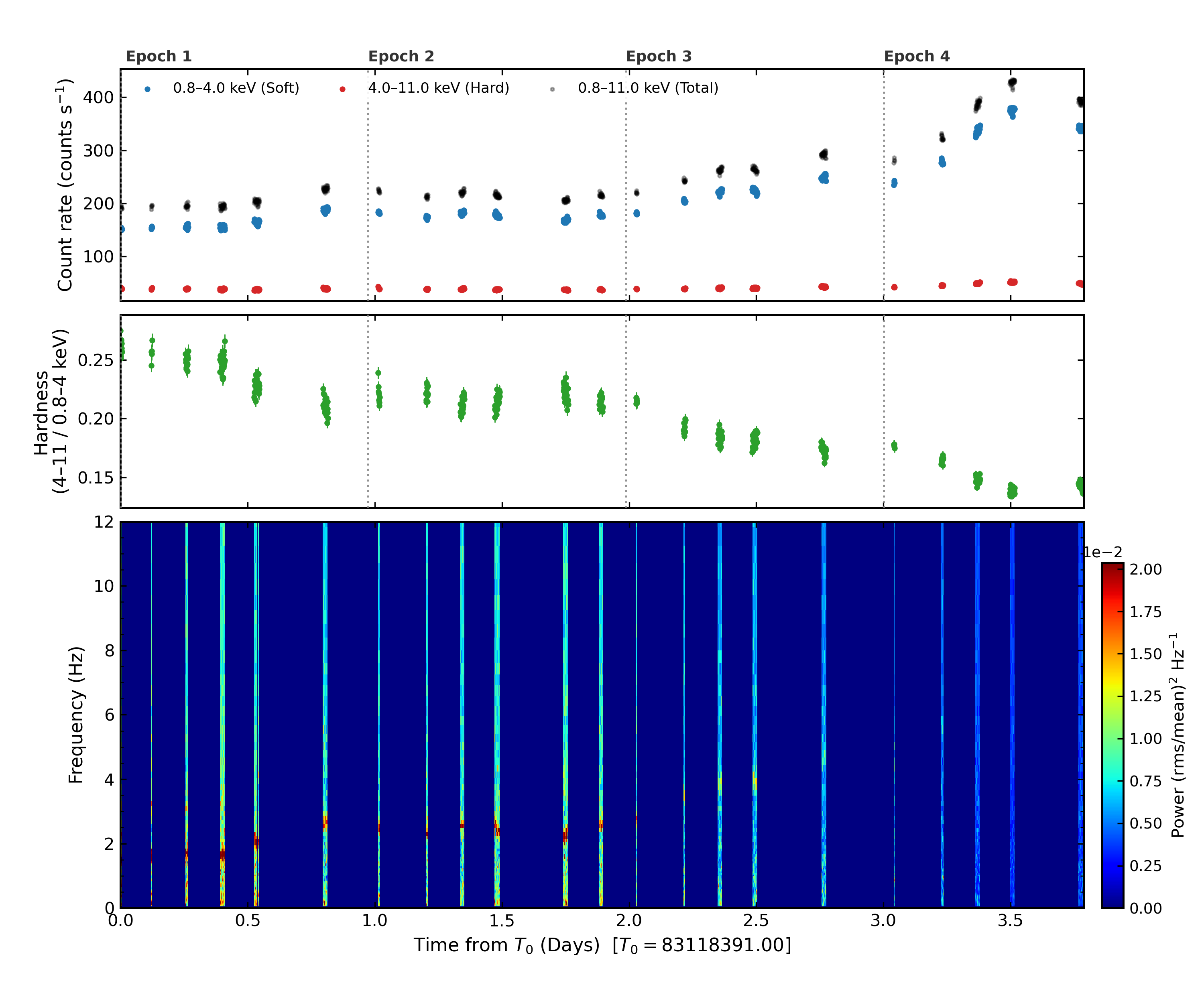}
   \caption{XSPECT timing properties of the BH X-ray binary MAXI J1750$-$327 during the XPoSat observation campaign across epochs 1–4. The top panel displays the light curves in the 0.8–4.0 keV (blue) and 4.0–11.0 keV (red) energy bands binned at 64 s, along with the total count rate in the 0.8–11.0 keV band (black). The middle panel presents the corresponding hardness ratio, defined as the ratio of hard to soft count rates (4.0–11.0 keV / 0.8–4.0 keV). The bottom panel shows the dynamic power spectrum across epochs 1–-4 generated with segment lengths of 128 s (sub-segment averaging of 16 s) and normalized to fractional rms-squared units [$(\text{rms}/\text{mean})^2\text{ Hz}^{-1}$], spanning the 0.05–12.0 Hz frequency range. An upward-migrating type-C QPO (evolving from $\sim 1.65\text{ Hz}$ to $\sim 8.65\text{ Hz}$) and its correlated band-limited noise are prominently observed before both features quench simultaneously toward the end of epoch 4.}
   \label{Fig:dynamic_ps}
\end{figure*}

\begin{figure*}[!t]
    \centering
    \includegraphics[width=0.6\textwidth]{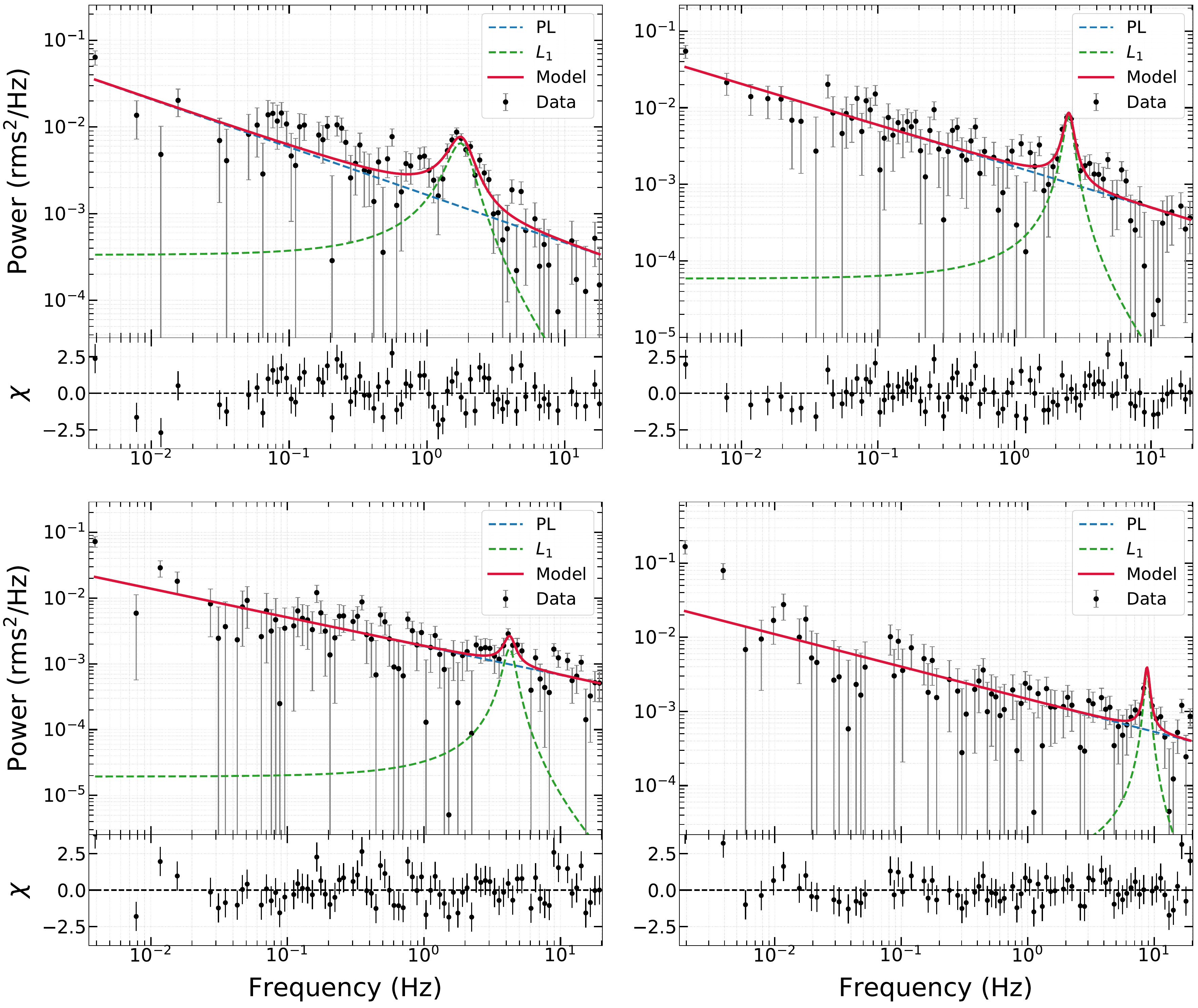}
    \caption{Poisson-noise-subtracted power density spectra (PDS) of MAXI J1750$-$327 obtained with XSPECT during epochs~1--4 with QPO detections: epoch~1 (top left), epoch~2 (top right), epoch~3 (bottom left), and epoch~4 (bottom right). Solid red curves show the best-fitting total models, while blue and green dashed lines represent the power-law ($PL$) and QPO Lorentzian ($L_1$) components, respectively. 
    }
    \label{Fig:PDS}
\end{figure*}


\section{Results}
\subsection{Timing Analysis} \label{sec:timing_analysis}

We present the daily long-term \textit{MAXI}/GSC light curve of MAXI~J1750$-$327 during its first-ever detected outburst in 2026 across multiple energy bands, alongside the corresponding hardness ratio (HR) in Figure~\ref{Fig:maxi_lc}. The specific observation epochs targeted by our \textit{XSPECT} pointing campaign are highlighted by vertical dashed lines. In the \textit{MAXI} hardness--intensity diagram (HID; Figure~\ref{Fig:HID}), the 2--20~keV flux is plotted as a function of  HR  (4--10~keV)/(2--4~keV), where the source traces a clear counter-clockwise ``q''-shaped pattern typical of BH X-ray binary outbursts \citep{2000arxt.confE..22H}.

We also present the HID corresponding to the 11 \textit{XSPECT} observation epochs in Figure~\ref{Fig:XSPECT_HID}, with the color scale tracing the chronological evolution of the source. The trajectory in the HID demonstrates a prominent spectral evolution that aligns consistently with the behavior captured by \textit{MAXI}/GSC. Notably, both the \textit{MAXI}/GSC and \textit{XSPECT} datasets indicate that a rapid spectral state transition takes place between the third and fourth observation epochs (around MJD~61274--61275), marked by a sharp drop in spectral hardness (Figure \ref{Fig:maxi_lc}e) accompanied by a substantial increase in flux (Figure \ref{Fig:maxi_lc}a,b,c ). Following this transition, the source enters a softer state where subsequent observations cluster within a relatively localized region in the HID, showing moderate intensity fluctuations at low HRs.

The light curves for epochs~1--4 across different energy bands of XSPECT are shown in the upper panel of Figure.~\ref{Fig:dynamic_ps}, binned at $64\rm\,s$. In the middle panel, we show the HR, defined as the ratio of count rates in the hard ($4.0$--$11.0\rm\,keV$) to soft ($0.8$--$4.0\rm\,keV$) bands, using the same time binning. Crucially, inspection of these upper two panels reveals a steady increase in the low-energy count rate accompanied by a systematic softening of the HR from epoch~1 to~4, clearly pointing to a spectral transition from the hard to the soft state. To track the timing evolution across these first four observations, we constructed dynamic power density spectra (DPDS; lower panel). Specifically, we segmented the light curves into $128\rm\,s$ intervals, applied Welch's method using $16\rm\,s$ sub-segments to suppress Poisson noise, and geometrically rebinned the resulting spectra by a factor of $1.05$ over the $0.05$--$12\rm\,Hz$ frequency range. We focused this analysis on these four epochs because they exhibit a prominent QPO that drifts from $\sim$$1\rm\,Hz$ to $\sim$$9\rm\,Hz$, consistent with the expected timing signatures during such a transition. A comprehensive analysis and modeling of this QPO feature is presented in Section~\ref{sec:timing_analysis}.


We investigated the X-ray aperiodic variability by computing Fourier power density spectra (PDS) from continuous 256s segments of the screened light curves spanning the $\sim 0.004$-$20$\,Hz frequency window for the initial three epochs, and from 512s segments for epoch 4 to better resolve a low-frequency feature. The white noise level arising from Poisson statistics was subsequently subtracted from each generated PDS. We rebinned the PDS with a factor of 1.08 in a geometrical series manner. The resulting power spectra were scaled according to the Miyamoto convention \citep{1991ApJ...383..784M, 1997scma.conf..321V}, resulting in power density in fractional rms units of $\mathrm{rms}^2/\mathrm{Hz}$. The 0.025 s light curves allowed the PDS up to Nyquist frequency of 20 Hz. 

The PDS generated for each day is modeled using Lorentzian and power-law functions. The power-law model ($A\nu^{-\alpha}$) with normalization $A$ and index $\alpha$ is used to model the low-frequency band-limited noise (BLN), alongside one standard Lorentzian profiles ($L_1$) to describe peaked QPO features. The Lorentzian component is parameterized as:
\begin{equation}
L(\nu) = \frac{K}{\pi} \frac{\mathrm{FWHM} / 2}{(\nu - \nu_{\mathrm{QPO}})^2 + (\mathrm{FWHM} / 2)^2},
\end{equation}
where $\nu_{\mathrm{QPO}}$ is the centroid frequency, FWHM is the full width at half-maximum, and $K$ denotes the total integrated power. The discrete QPO profiles are characterized using this profile, with the fractional rms amplitude derived directly from the square root of the normalization ($\mathrm{rms} = \sqrt{K} \times 100\%$). 

Representative PDSs for epochs 1$-$4 are shown in Figure \ref{Fig:PDS}. The resulting best-fit parameters of the lorentzians, characteristic quality factors ($Q = \nu_{\mathrm{QPO}} / \Delta\nu$), and fractional root mean square (rms) variability amplitudes are summarized in Table \ref{Table:QPO}.

During epoch 1, the PDS displays a relatively well-defined low-frequency QPO ($L_1$), superposed on a broader underlying noise excess. In epoch 2, the oscillation hardens and sharpens considerably, shifting to a higher frequency with an increased Q factor. The corresponding dynamic power spectra (Figure. \ref{Fig:dynamic_ps}) clearly track this upward-migrating type-C QPO, where red marks (using a power definition appropriate for visualizing this feature) distinctly highlight the feature's frequency evolution from $\sim 1.7$\,Hz to $\sim 2.5$\,Hz and its prominent correlation with the observed hard band-limited noise across these first two epochs. As the system continued its softening trajectory, epoch 3 showed a marked broadening of the timing structures, with the primary peaked component shifting further high in frequency. This broadening is evident in the dynamic power spectra (Figure. \ref{Fig:dynamic_ps}), which also registers a noticeable transition in the QPO's coherence and amplitude near the start of the third observation, indicative of a shift from type-C to type-B, just before the timing features significantly decrease in power. In epoch~4, utilizing the longer 512\,s segments, an oscillation re-establishes itself at a substantially higher centroid frequency, which we have fitted with a standard Lorentzian component (Figure.~\ref{Fig:PDS}). While this feature is less prominent in the dynamic power spectra (Figure.~\ref{Fig:dynamic_ps}) due to the 64,s binning used for that visualization, its presence in the integrated PDS over identical 512,s segments yields a detection significance of $3.29\sigma$.

We note that a consistent report on the detection of the $\sim$ 8 Hz QPO in this source during a similar epoch, as well as indications of QPO transitions, has also been made by the Insight-HXMT team \citep{2026ATel18026....1C, 2026ATel18024....1C}. Furthermore, our obtained QPO centroid frequencies are in broad agreement with the findings reported in \cite{2026ATel18010....1C}. In general, the fundamental centroid frequency shows a clear upward shift over consecutive days, while the fractional rms amplitude exhibits a decreasing trend, as detailed in Table \ref{Table:QPO}.


\begin{table}[!htbp]
\centering
\small
\caption{Best-fit timing parameters of the Lorentzian components for MAXI~J1750$-$327 obtained across epochs~1$-$4.}
\label{tab:qpo_parameters}
\setlength{\tabcolsep}{3pt}
\resizebox{\columnwidth}{!}{%
\begin{tabular}{lcccc}
\hline
\hline
\noalign{\smallskip}
\textbf{Parameter} & \textbf{Epoch 1} & \textbf{Epoch 2} & \textbf{Epoch 3} & \textbf{Epoch 4} \\
\noalign{\smallskip}
\hline
\noalign{\smallskip}
Component & $L_1$ & $L_1$ & $L_1$ & $L_1$ \\
$\nu_{\mathrm{QPO}}$ (Hz) & $1.75 \pm 0.05$ & $2.50 \pm 0.02$ & $4.23 \pm 0.16$ & $8.77 \pm 0.22$ \\
$\Delta\nu$ (Hz) & $0.82 \pm 0.16$ & $0.44 \pm 0.08$ & $0.91 \pm 0.52$ & $0.88 \pm 1.32$ \\
$Q$ & $2.1 \pm 0.4$ & $5.6 \pm 1.0$ & $4.63 \pm 2.64$ & $10.0 \pm 15.3$ \\
rms (\%) & $9.1 \pm 0.7$ & $7.2 \pm 0.4$ & $4.91 \pm 1.05$ & $6.8 \pm 2.4$ \\
$\chi^2 / \text{dof}$ & 1.56 & 1.10 & 1.35 & 1.19 \\

\noalign{\smallskip}
\hline
\end{tabular}%
}
\vskip 4pt
\parbox{\columnwidth}{%
\footnotesize
$L_{1}$ : fundamental QPO Lorentzian component

$\nu_{\mathrm{QPO}}$ : centroid frequency

$\Delta\nu$ : full width at half maximum

$Q = \nu_{\mathrm{QPO}} / \Delta\nu$ : quality factor

$\text{rms}$ : fractional root-mean-square amplitude

$\chi^2/\text{dof}$ : total fit chi-square per degrees of freedom .
}
\label{Table:QPO}
\end{table}

\begin{figure}[h!]
    \centering
    \includegraphics[width=1.0\linewidth]{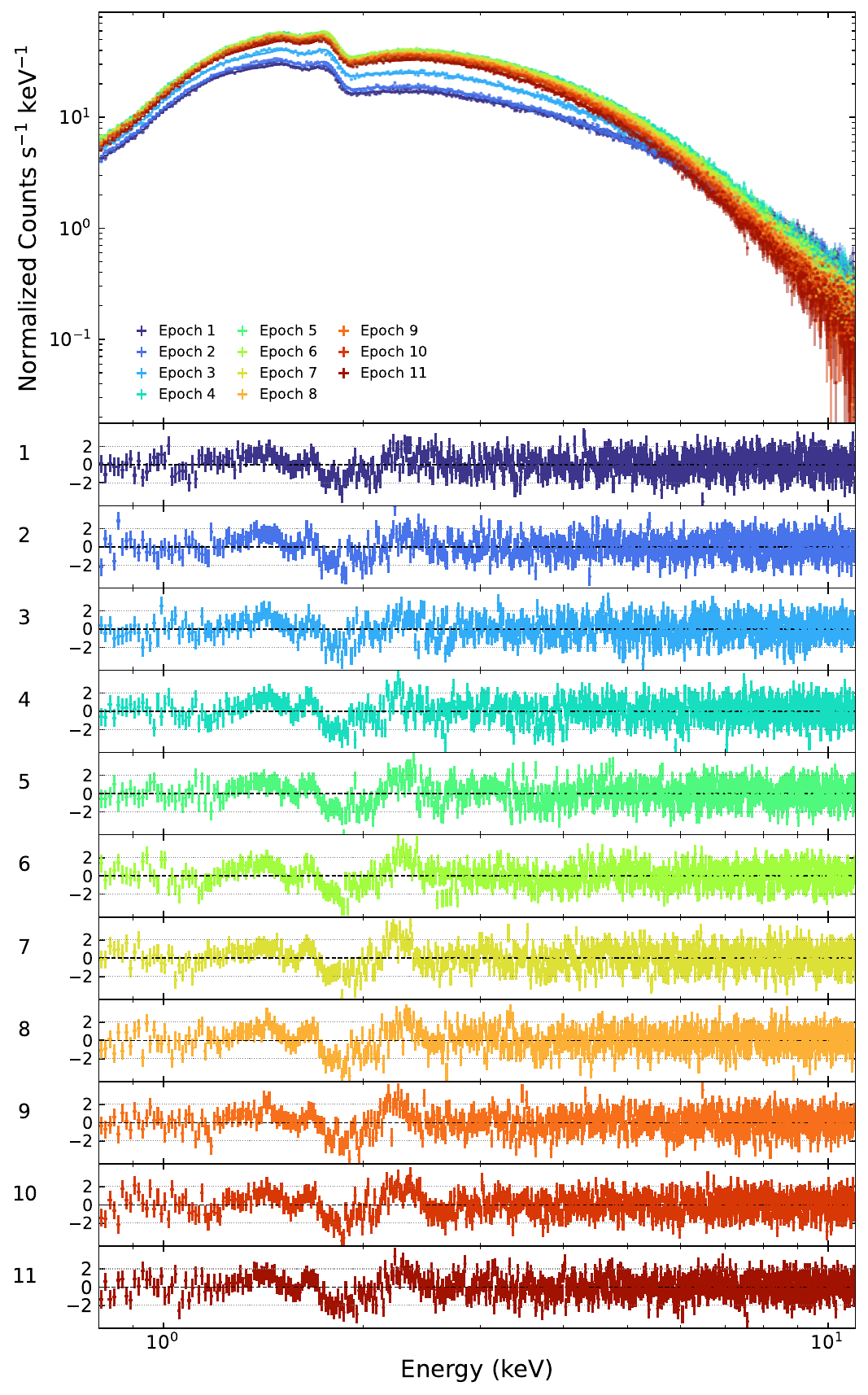}
    \caption{Day-wise energy spectra fitted in the 0.8$-$11.0\,keV energy band for epochs~1 to 11 using the Model \texttt{TBabs*(nthComp + gaussian + diskbb)}, incorporating an additional Gaussian component to account for the Fe~K$\alpha$ emission feature (except for epochs~3 and 11). The top panel displays the unfolded spectra along with the best-fit model components, whereas the lower stacked sub-panels show the corresponding fit residuals ($\Delta\chi$) for each epoch. The dashed horizontal line indicates zero residual, and the dotted lines denote the $\pm2\sigma$ deviation levels.}
    \label{Fig:delchi_Fe}
\end{figure}


\begin{table*}[!htbp]
\centering
\small
\caption{Best-fit spectral parameters for the Model - 2 : $TBABS^{*}(NTHCOMP + GAUSSIAN + DISKBB)$. }
\vspace{0.2cm} 
\label{tab:spectral_fits}
\renewcommand{\arraystretch}{1.3}
\resizebox{\textwidth}{!}{
\begin{tabular}{lcccccccccccc}
\hline
Epoch & $N_{\text{H}}$ & $\Gamma$ & $N_{\mathrm{nthComp}}$ & $E_{\text{line}}$ & $\sigma_{\text{line}}$ & $N_{\mathrm{Ga}}$ & $T_{\text{in}}$ & $N_{\mathrm{diskbb}}$ & $F_{\mathrm{nthComp}}$ &
$F_{\mathrm{diskbb}}$ & $F_{\mathrm{total}}$ & $\chi^2 / \text{dof}$ \\
& ($10^{22}\text{ cm}^{-2}$) & & & (keV) & (keV) & ($10^{-2}$) & (keV) & & \multicolumn{3}{c}{($10^{-8}\ \mathrm{erg\ cm^{-2}\ s^{-1}}$, 0.8-10.0 keV)} \\
\hline
1 & $1.18_{-0.04}^{+0.05}$ & $2.03_{-0.02}^{+0.02}$ & $9.68_{-0.83}^{+0.93}$ & $6.70_{-0.09}^{+0.07}$ & $0.13_{-0.06}^{+0.09}$ & $0.92_{-0.35}^{+0.39}$ & $0.43_{-0.04}^{+0.04}$ & $11861_{-3402}^{+6305}$ & $5.16^{+0.01}_{-0.01}$ & $0.43_{-0.001}^{+0.001}$ & $5.60^{+0.01}_{-0.02}$ & $880.45 / 807$ \\
2 & $1.14_{-0.02}^{+0.02}$ & $2.10_{-0.02}^{+0.02}$ & $8.99_{-0.71}^{+0.82}$ & $6.54_{-0.04}^{+0.03}$ & $0.006_{-0.006}^{+0.075}$ & $0.004<$ & $0.52_{-0.03}^{+0.03}$ & $6716_{-1032}^{+1128}$ & $5.07^{+0.01}_{-0.01}$ & $0.65^{+0.01}_{-0.02}$ & $5.73^{+0.01}_{-0.02}$ & $863.01/810$\\
3 & $1.15_{-0.02}^{+0.02}$ & $2.38_{-0.05}^{+0.04}$ & $10.36_{-1.10}^{+1.31}$ & ------ & ------ & ------ & $0.69_{-0.04}^{+0.04}$ & $3041_{-376}^{+303}$ & $5.98^{+0.02}_{-0.02}$ & $1.01^{+0.01}_{-0.01}$ & $6.99^{+0.02}_{-0.02}$ & $933.65 / 810$ \\
4 & $1.19_{-0.01}^{+0.01}$ & $2.55_{-0.18}^{+0.17}$ & $6.78_{-1.33}^{+1.56}$ & $6.74_{-0.05}^{+0.05}$ & $0.11_{-0.05}^{+0.08}$ & $1.41_{-0.44}^{+0.52}$ & $1.01_{-0.02}^{+0.02}$ & $2744_{-163}^{+123}$   & $5.02^{+0.02}_{-0.02}$ & $4.87^{+0.01}_{-0.02}$ & $9.92^{+0.02}_{-0.02}$ & $938.77 / 807$  \\
5 & $1.18_{-0.01}^{+0.01}$ & $2.57_{-0.12}^{+0.11}$ & $8.32_{-1.26}^{+1.41}$ & $6.60_{-0.07}^{+0.08}$ & $0.16_{-0.08}^{+0.10}$ & $1.32_{-0.49}^{+0.52}$ & $0.92_{-0.03}^{+0.02}$ & $2795_{-190}^{+150}$   & $5.62^{+0.02}_{-0.02}$ & $3.28^{+0.01}_{-0.02}$ & $8.92^{+0.01}_{-0.02}$ & $945.40 / 807$  \\
6 & $1.22_{-0.01}^{+0.01}$ & $2.68_{-0.19}^{+0.17}$ & $7.06_{-1.43}^{+1.63}$ & $6.78_{-0.11}^{+0.09}$ & $0.19_{-0.11}^{+0.16}$ & $1.23_{-0.53}^{+0.74}$ & $0.99_{-0.02}^{+0.02}$ & $2994_{-186}^{+138}$   & $4.94^{+0.02}_{-0.01}$ & $4.93^{+0.02}_{-0.02}$ & $9.89^{+0.02}_{-0.02}$ & $840.46 / 807$  \\
7 & $1.23_{-0.01}^{+0.01}$ & $2.65_{-0.20}^{+0.19}$ & $6.11_{-1.31}^{+1.49}$ & $6.61_{-0.09}^{+0.10}$ & $0.21_{-0.08}^{+0.10}$ & $1.29_{-0.48}^{+0.57}$ & $0.99_{-0.02}^{+0.02}$ & $3144_{-164}^{+124}$   & $4.32^{+0.02}_{-0.02}$ & $5.19^{+0.02}_{-0.02}$ & $9.53^{+0.01}_{-0.02}$ & $962.82 / 807$  \\
8 & $1.20_{-0.01}^{+0.01}$ & $2.57_{-0.22}^{+0.22}$ & $5.12_{-1.19}^{+1.51}$ & $6.61_{-0.13}^{+0.18}$ & $0.20_{-0.19}^{+0.16}$ & $1.03_{-0.49}^{+0.63}$ & $0.99_{-0.02}^{+0.02}$ & $3142_{-155}^{+113}$   & $3.71^{+0.02}_{-0.01}$ & $5.22^{+0.02}_{-0.02}$ & $8.95^{+0.02}_{-0.02}$ & $930.44 / 807$  \\
9 & $1.21_{-0.02}^{+0.01}$ & $2.39_{-0.55}^{+0.30}$ & $3.73_{-1.79}^{+1.63}$ & $7.06_{-0.35}^{+0.26}$ & $0.80_{-0.24}^{+0.50}$ & $3.55_{-1.62}^{+1.60}$ & $1.00_{-0.03}^{+0.04}$ & $3355_{-116}^{+104}$   & $2.90^{+0.02}_{-0.01}$ & $5.84^{+0.01}_{-0.02}$ & $8.79^{+0.01}_{-0.02}$ & $876.91 / 807$ \\
10 & $1.18_{-0.01}^{+0.01}$ & $2.36_{-0.28}^{+0.27}$ & $3.38_{-0.99}^{+1.24}$ & $6.72_{-0.08}^{+0.07}$ & $0.20_{-0.08}^{+0.11}$ & $1.61_{-0.54}^{+0.65}$ & $1.00_{-0.02}^{+0.02}$ & $3317_{-103}^{+95}$    & $2.65^{+0.02}_{-0.02}$ & $5.66^{+0.02}_{-0.02}$ & $8.33^{+0.02}_{-0.02}$ & $832.13 / 807$ \\
11 & $1.22_{-0.01}^{+0.01}$ & $2.93_{-0.23}^{+0.22}$ & $5.82_{-1.31}^{+1.59}$ & ----- & ----- & ----- & $0.93_{-0.01}^{+0.01}$ & $3388_{-236}^{+169}$  & $3.57^{+0.02}_{-0.02}$ & $4.29^{+0.01}_{-0.01}$ & $7.87^{+0.01}_{-0.01}$ & $859.14 / 810$  \\
\hline
\end{tabular}
}

\parbox{\textwidth}{%
\footnotesize
Errors are quoted at the 90\% confidence level ($\Delta\chi^2 = 2.706$, $1.6\sigma$). 

$N_{\mathrm{H}}$ is the hydrogen column density in $10^{22}\ \mathrm{cm}^{-2}$.

$\Gamma$ and $N_{\mathrm{nthComp}}$ represent the non-thermal photon index and normalization of the \textsc{nthomp} component, respectively. The hot electron plasma temperature ($kT_e$) is fixed at 50 keV. 

$T_{\mathrm{in}}$ and $N_{\mathrm{diskbb}}$ are the inner disk temperature (in keV) and normalization of \textsc{diskbb} respectively.

$E_{line}$, $\sigma_{line}$, and $N_{Ga}$ are the Gaussian component to incorporate the Fe $K\alpha$ emission line in the spectrum, the width of the line (both in keV) and the Gaussian normalization. The inclusion of the Gaussian Fe~K$\alpha$ component resulted an $F$-test detection significance above $3\sigma$ ($>99.9\%$) across most epochs, except for epoch~2 and 11 where the feature was detected marginally at $2.9\sigma$, while for epochs~3 and 11 the line addition was statistically insignificant.

$F_{\mathrm{nthComp}}$, $F_{\mathrm{diskbb}}$, and $F_{\mathrm{total}}$ represent the unabsorbed non-thermal nthcomp flux, disk flux, and total fluxes in the 0.8-10.0\,keV range.

$\chi^2/\text{dof}$ : total fit chi-square per degrees of freedom


}
\label{tab:spec_table}
\end{table*}


\begin{figure}
    \centering
    \includegraphics[width=1.0\linewidth]{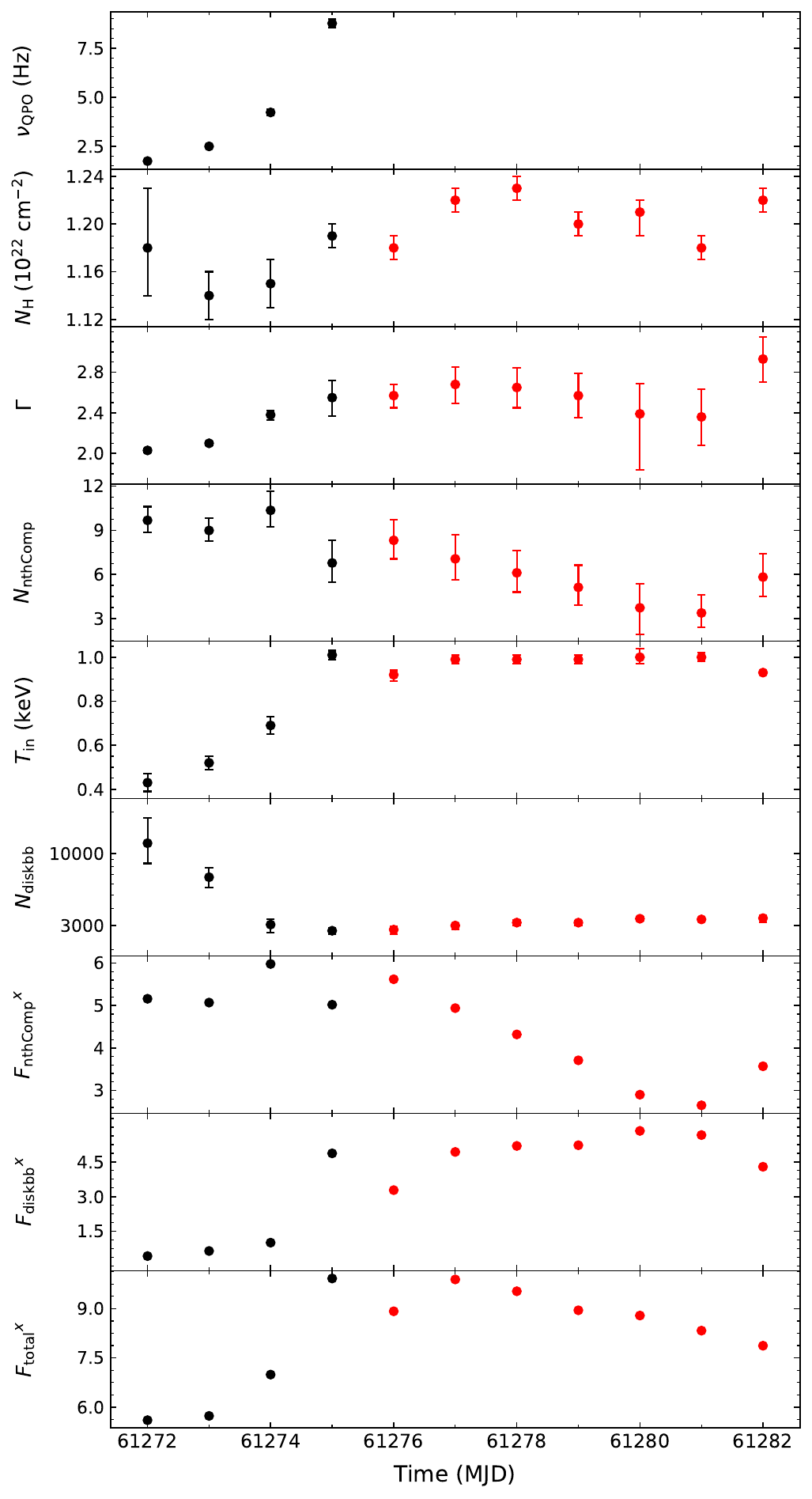}
    \caption{Day-wise evolution of the best-fit spectral parameters for MAXI~J1750$-$327 obtained using Model 1 : \texttt{TBabs*(nthComp + gaussian + diskbb)} during epoch 1 $-$ 11. The black circles denote epochs where the QPO is present, while the red circles represent the epochs without QPO detection. First panel : QPO centroid frequency ($\nu_{\mathrm{QPO}}$ in Hz), second panel : hydrogen column density ($N_{\mathrm{H}}$ in units of $10^{22}\ \mathrm{cm}^{-2}$), third panel : photon index ($\Gamma$), fourth panel : normalization of the nthComp component ($N_{\mathrm{nthComp}}$), fifth panel : inner disk temperature ($T_{\mathrm{in}}$ in keV), sixth panel : normalization of the disk blackbody ($N_{\mathrm{diskbb}}$, plotted on a logarithmic scale). Seventh panel : the unabsorbed non-thermal flux, eighth panel : unabsorbed thermal flux, ninth panel : unabsorbed total flux in the $0.8\text{-}11.0\ \mathrm{keV}$ energy range in the unit (x $=10^{-8}\ \mathrm{erg\ cm^{-2}\ s^{-1}}$). }
    \label{Fig:spec_evo}
\end{figure}


\subsection{Spectral analysis \label{sec:spec_analysis}} 
We performed spectral analysis of MAXI J1750$-$327 across the 11 continuous observation epochs (MJD 61272-61282) using \textit{XSPECT} data in the 0.8-11.0 keV band. Photons with energies above 11 keV were excluded due to significant background contamination in this range. A systematic uncertainty of 3\% was adopted in the channels 96--148 (corresponding to $\sim 1.2$--$1.8\text{ keV}$) throughout the analysis, as recommended by the instrument team \citep{2025JATIS..11d4007C}. 
We used wilms abundances \citep{2000ApJ...542..914W} with cross-section of Verner \citep{1996ApJ...465..487V} for the absorption.
Interstellar absorption along the line of sight was modeled across all spectra using \texttt{TBabs} with standard interstellar abundances and cross-sections \citep{2000ApJ...542..914W}. 

To characterize the underlying emission, we initially applied a two-component baseline model comprising a multi-color disk blackbody (diskbb \cite{1984PASJ...36..741M}, \cite{1986ApJ...308..635M}) and a comptonization model (nthcomp \cite{1999MNRAS.309..561Z}). In this continuum framework, the seed photon temperature ($kT_{\mathrm{bb}}$) for Comptonization was tied to the inner disk temperature ($T_{\mathrm{in}}$), and the electron temperature of the hot corona ($kT_e$) was fixed at 50 keV, following typical conventions in X-ray binary modeling (e.g., \cite{2024MNRAS.532.4486A}) when high-energy spectral rollover cannot be constrained within the \textit{XSPECT} operating passband. Although this continuum configuration described the overall spectral curvature reasonably well, inspecting the fit residuals revealed localized positive excesses in the $\sim$ 6.0–7.0 keV range across multiple epochs, indicative of iron K$\alpha$ reflection features. To properly account for this emission feature, we incorporated an additional Gaussian line component (\texttt{gaussian}), formalizing the total spectral model as \texttt{TBabs*(nthComp + gaussian + diskbb)}. An $F$-test confirmed that the inclusion of the Gaussian profile significantly improved the fits beyond the $3\sigma$ confidence level ($>99.9\%$) for most observations, with the exception of epochs~2 and 11 where the line was only marginally detected or statistically insignificant. The line centroid varied between $E_{\mathrm{line}} \approx 6.54$ keV and $7.06$ keV with intrinsic widths of $\sigma_{\mathrm{line}} \sim 0.11$-$0.80$ keV, confirming a persistent, moderately broadened iron line. Additionally, minor residual structures were noticed at lower energies, notably below $\sim$2.0 keV and around $\sim$2.2 keV.  Instrument-related residuals in this soft energy regime, particularly associated with detector Al and Si K-absorption edges ($\sim$1.5 and $\sim$1.8 keV), have also been documented in earlier XSPECT observations \citep{2026arXiv260721365N}. Despite incorporating the recommended systematic uncertainty in this band, subtle residuals persisted. However, we found that attempting to account for these remaining features with ad hoc components (such as additional Gaussians or edges) introduced strong parameter degeneracies with the continuum. Therefore, to preserve the physical consistency and reliability of the continuum fit parameters, no extra empirical components were included at these low energies. The day-wise unfolded energy spectra, best-fit composite models and corresponding stacked residuals across all epochs are shown in Figure~\ref{Fig:delchi_Fe}, while the derived best-fit parameters, fit statistics, and unabsorbed component fluxes are summarized in Table~\ref{tab:spectral_fits}. Although the convolution model \texttt{thcomp} was initially tested, its multiplicative coupling introduced strong parameter degeneracies with the continuum and caused unphysical variations in $N_{\mathrm{H}}$. Consequently, we adopted the additive Comptonization model \texttt{nthComp} to ensure spectral stability and break these degeneracies throughout our fits.

The temporal evolution of the best-fit spectral parameters along with the timing properties is shown in Figure~\ref{Fig:spec_evo}. Throughout the entire observation period, the column density remained remarkably steady at $N_{\mathrm{H}} \approx (1.14$-$1.23) \times 10^{22} \mathrm{cm}^{-2}$, confirming minimal variability in the line-of-sight absorbing medium, shows excellent agreement with independent measurements from instruments featuring sensitive low-energy coverage such as EP-WXT and \textit{Insight}-HXMT \citep{2026ATel18026....1C}.  The emission parameters showed marked day-to-day spectral softening:
During the first three epochs (epochs 1$-$3), the thermal disk temperature, $T_{\mathrm{in}}$ progressively increased from $0.43 \pm 0.04$\,keV to $0.69 \pm 0.04$\,keV, while the disk normalization $N_{\mathrm{diskbb}}$ dropped sharply from $11861_{-3402}^{+6305}$ to $3041_{-376}^{+303}$. In epoch~4, the disk heated rapidly to $T_{\mathrm{in}} \approx 1.01 \pm 0.02$ keV, stabilizing around $\sim 0.92$-$1.00$ keV for the remainder of the outburst, accompanied by a stable normalization ($N_{\mathrm{diskbb}} \approx 2744$-$3388$). This thermal evolution is broadly consistent with the findings of \citet{2026ATel18010....1C}, who similarly reported an initial disk temperature of $T_{\mathrm{in}} \sim 0.5$\,keV during epochs~1--2 that subsequently climbed to $\sim 1$\,keV by epoch~4.
The non-thermal photon index softened substantially over time, steepening from $\Gamma = 2.03 \pm 0.02$ in epoch~1 to $\Gamma = 2.55_{-0.18}^{+0.17}$ in epoch~4 and ultimately reaching $\Gamma = 2.93_{-0.23}^{+0.22}$ by epoch~11. 
The Comptonization normalization decreased from $N_{\mathrm{nthComp}} \approx 9.68$ to $3.38$ (with a slight increase to $5.82$ in epoch~11), reflecting an overall reduction in the non-thermal scattering efficiency of the corona.
The unabsorbed flux distribution exhibits a clear transition from non-thermal to thermal dominance. 

This quantitative restructuring of the energy spectrum characterized by the steepening of $\Gamma$, the rise of $T_{\mathrm{in}}$ above $\sim 0.9$\,keV, and the shift to disk-dominated flux correlates closely with the disappearance of coherent QPOs after epoch~4, tracking the state transition of MAXI~J1750$-$327 from the hard-intermediate state to the soft-intermediate/soft state.


\begin{figure*}[!t]
    \centering
    \includegraphics[width=1.0\textwidth]{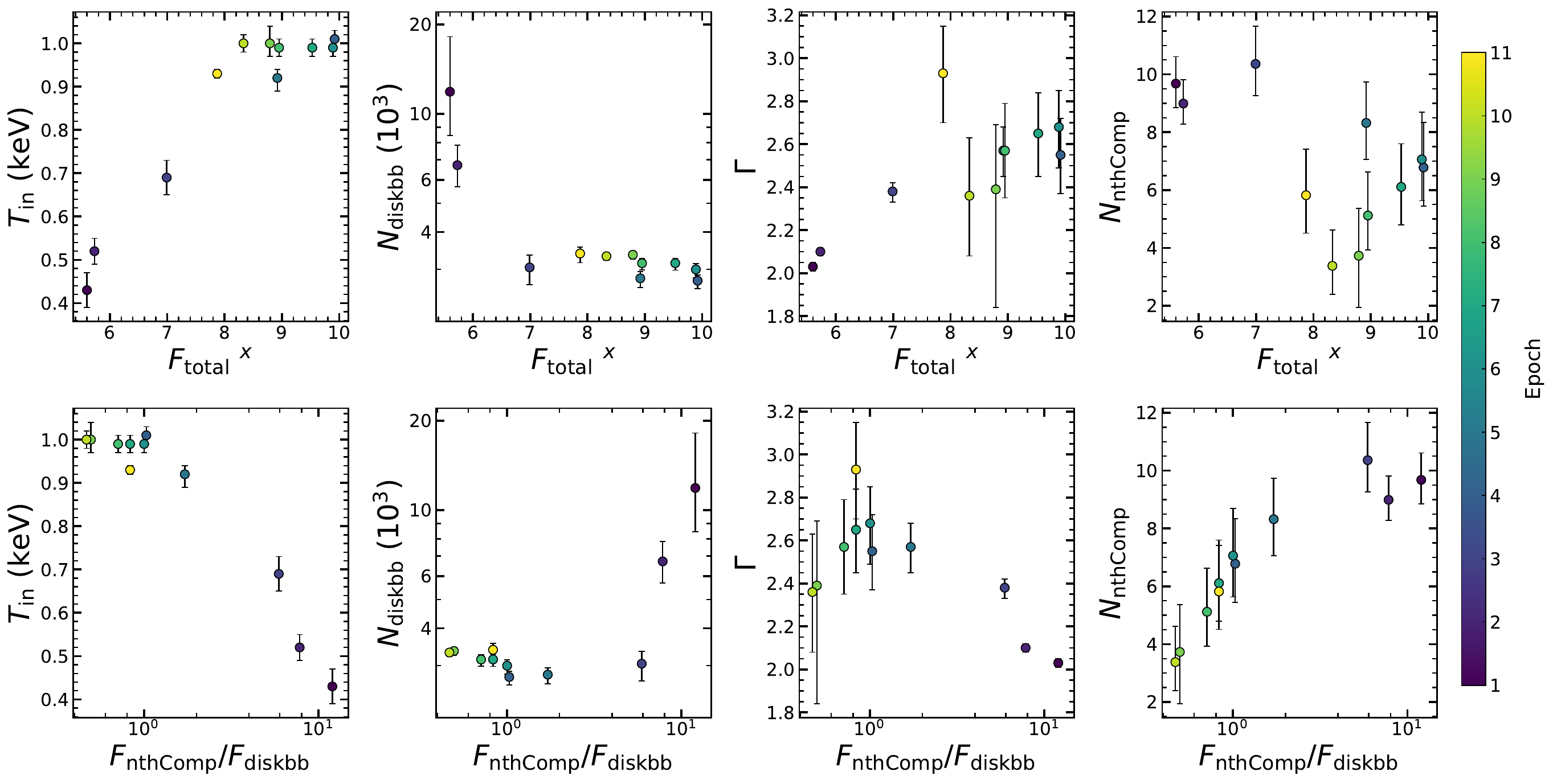}
    \caption{Logarithmic evolution of the spectral parameters across epochs 1--11 as a function of total unabsorbed flux $F_{\mathrm{total}}$ in the unit (x $=10^{-8}\ \mathrm{erg\ cm^{-2}\ s^{-1}}$) (upper panels) and the non-thermal to thermal flux ratio $F_{\mathrm{nthComp}}/F_{\mathrm{diskbb}}$ (lower panels). From left to right: inner disk temperature ($T_{\mathrm{in}}$), disk normalization ($N_{\mathrm{diskbb}}$), photon index ($\Gamma$), and nthComp normalization ($N_{\mathrm{nthComp}}$). Data points are color-coded according to the observation epoch, indicated by the color bar on the right.}
    \label{Fig:corr_spec}
\end{figure*}


\begin{figure*}[!t]
    \centering
    \includegraphics[width=0.82\textwidth]{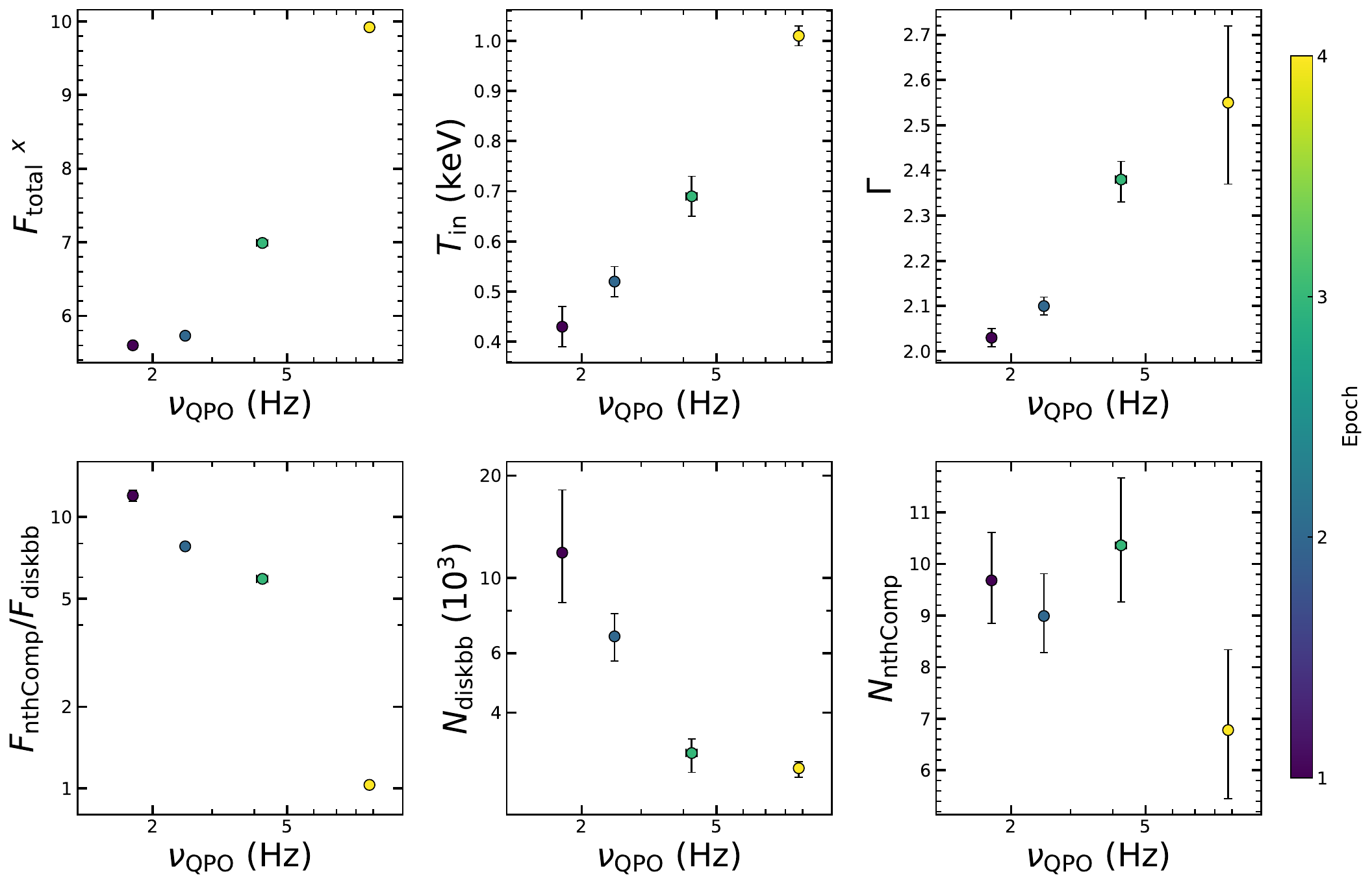}
    \caption{Logarithmic evolution of the spectral parameters with the QPO centroid frequency ($\nu_{\mathrm{QPO}}$) for epochs 1-4: total flux $F_{\mathrm{total}}$ in the unit (x $=10^{-8}\ \mathrm{erg\ cm^{-2}\ s^{-1}}$) (top left), inner disk temperature $T_{\mathrm{in}}$ (top middle), photon index $\Gamma$ (top right), ratio of the fluxes $F_{\mathrm{nthComp}}/F_{\mathrm{diskbb}}$ (bottom left), disk normalization $N_{\mathrm{diskbb}}$ (bottom middle),  and Comptonization normalization $N_{\mathrm{nthComp}}$ (bottom right). Data points are color-coded according to the observation epoch, indicated by the color bar on the right.}
    \label{fig:spec_time_corr}
\end{figure*}

\section{Discussion}
In this work, we present a comprehensive spectral and timing analysis of MAXI~J1750$-$327 during its first-ever detected outburst in August 2026. 
Several missions including \textit{Einstein Probe} (EP-FXT/WXT), \textit{Insight}-HXMT, \textit{NuSTAR}, and \textit{XPoSat} collected X-ray data during this period. In this study, we utilize XSPECT observations to carry out our detailed spectro-temporal analysis.
The MAXI light curve (Figure.~\ref{Fig:maxi_lc}) shows that the flux above $10\text{ keV}$ decreases systematically during the XSPECT observations, whereas the flux below $10\text{ keV}$ increases across the first four epochs before stabilizing. The hardness--intensity diagrams (Figures.~\ref{Fig:HID} and \ref{Fig:XSPECT_HID}) mirror this behavior, indicating that softer photons gradually dominate the emission. Crucially, both datasets indicates a rapid spectral state transition between the third and fourth observation epochs ($\sim\text{MJD } 61274\text{--}61275$), where the spectral hardness drops sharply while the soft energy band flux increases substantially, leading the source into a softer state that clusters within a localized region of the HID (Figures.~\ref{Fig:HID} and \ref{Fig:XSPECT_HID}). Focusing on the dynamic power spectra across the first four epochs (Figure.~\ref{Fig:dynamic_ps}), 
we trace the evolution of the timing features and identify a distinct transition from a type-C to a type-B QPO between epochs 2 and 3, consistent with \citet{2026ATel18024....1C, 2026ATel18026....1C}. A low-frequency variability feature is present when the type-C QPO is present (Figure.~\ref{Fig:dynamic_ps}), and it disappears from the dynamic power spectra when the type-B QPO occurs. Similar observation has been seen for Swift J1727.8--1613\citep{2025A&A...699A...9J}.
Finally, energy-resolved light curves over these four days demonstrate a steady rise in low-energy counts accompanied by systematic spectral softening, directly establishing a transition from the hard to the soft state (upper panel of Figure.~\ref{Fig:dynamic_ps}).
The type-C QPO centroid frequency exhibits a systematic upward shift across epochs 1--2 (Figure.~\ref{Fig:dynamic_ps}), drifting from $1.75\pm0.05$ to $2.50\pm0.02$\,Hz. We classify this feature as a type-C QPO based on its relatively high fractional rms amplitude ($7.2\pm0.4\%$--$9.1\pm0.7\%$), moderate quality factor ($Q \approx 2.1\pm0.4$--$5.6\pm1.0$) \citep{2005ApJ...629..403C}.
The feature evolves into a sharper type-B QPO characterized by reduced rms amplitude ($\sim 4.9\%$--$6.8\%$) before disappearing entirely across the remaining observations (epochs 5--11) (Table~\ref{Table:QPO}).

To understand the physical changes inside the accretion flow that drive the QPO transition and turn off, we performed the spectral analysis and correlated the evolution with the QPO parameters. 
Tracking the evolution of these parameters across individual days in Figures  \ref{Fig:spec_evo} and \ref{Fig:corr_spec} reveals that most parameters change noticeably alongside the evolution of the QPO. We observed that following the disappearance of the QPO, the spectral parameters generally stabilize, except for the non-thermal flux and the nthComp normalization.
The inner disk temperature ($T_{\mathrm{in}}$) increases abruptly from $\sim 0.69\text{ keV}$ to $\sim1.01\text{ keV}$ during the first four observations accompanied by a steep drop in disk normalization ($N_{\mathrm{diskbb}}$) (Figure.\ref{Fig:corr_spec}) that signals the inward advance and stabilization of the apparent inner disk edge towards the ISCO \citep{1973A&A....24..337S} before settling into a steady range ($0.92$-$1.00\text{ keV}$) for epochs~5-11. 
Concurrently, the non-thermal Comptonized flux and normalization decline markedly as the photon index steepens ($\Gamma \sim 2.38 \rightarrow 2.93$), reflecting the progressive quenching and cooling of the corona. 
During this period, contemporaneous observations reported an upward-drifting QPO spanning $1.87$-$8.77\text{ Hz}$  with \textit{Insight}-HXMT confirming a prominent peak at $8.260 \pm 0.073\text{ Hz}$ on 23 August \citep{2026ATel18026....1C, 2026ATel18024....1C}. 

Across several BH LMXBs, the suppression of the QPO has often been observed alongside a stationary accretion disk. In GX~339$-$4 \citep{2009MNRAS.392..992D, 2021MNRAS.508..287S, 2023MNRAS.521.3570Y} and H~1743$-$322 \citep{2012A&A...541A...6S}, the inner disk radius remained locked during transitions, implying that QPO quenching was driven entirely by coronal evolution rather than changing disk truncation. Similar static-disk behavior has been reported in Swift J151857.0$-$572147 \citep{2027JHEAp..5600738R} and Swift~J1658.2$-$4242 \citep{2020A&A...641A.101B}. In sharp contrast to this static-disk scenario, MAXI J1750$-$327 exhibits an active, dynamical inward migration of the thin disk during the transition. The parameter trajectories presented in Figure \ref{Fig:corr_spec} and Figure \ref{fig:spec_time_corr} illustrate the tight coupling between the emergent accretion geometry and the QPO properties. 
We observed the correlations between the timing and spectral parameters (Figure \ref{fig:spec_time_corr}) which provide physical support for the Lense$-$Thirring precession framework, where the QPO frequency scales inversely with the truncation radius as $\nu \propto r_{\mathrm{in}}^{-3}$ \citep{2009MNRAS.397L.101I, 2019NewAR..8501524I}. We also observed that the  QPOs display a significant positive correlation with the total flux $F_{\mathrm{total}}$ (Figure.~\ref{fig:spec_time_corr}, top-left). Assuming that the mass accretion rate scales with the total flux, this result indicates a direct connection between $\mathrm{QPO}$ frequency and the instantaneous mass accretion rate of the system, matching the correlated frequency evolution reported for GRS~1915+105 by \cite{2021ApJ...909...63L}. Furthermore, $\mathrm{QPO}$ frequency exhibits a tight positive correlation with the non-thermal photon index $\Gamma$ (Figure.~\ref{fig:spec_time_corr}, top-right), consistent with the findings of \cite{2003A&A...397..729V}.
In this framework, the QPO frequency connects to a characteristic truncation radius within the accretion flow, with the emerging soft photons subsequently undergoing inverse-Compton scattering in the corona.

As the geometrically thin disk advances inward, it contracts the radial extent of the hot inner flow, driving the positive correlation between $\nu_{\mathrm{QPO}}$ and $T_{\mathrm{in}}$, the reduction in emitting area ($N_{\mathrm{diskbb}} \propto r_{\mathrm{in}}^2$; Figures \ref{fig:spec_time_corr}), and the rapid upward drift toward $\sim 8\text{ Hz}$. Simultaneously, the advancing disk intercepts a larger solid angle and injects abundant soft seed photons into the corona \citep{2007A&ARv..15....1D}, driving the steep inverse dependence of $\nu_{\mathrm{QPO}}$ on the flux ratio $F_{\mathrm{nthComp}}/F_{\mathrm{diskbb}}$ (Figure \ref{fig:spec_time_corr}). 


Once the inner disk tends to approach ISCO as the spectral state transition is taking place from the hard to the soft-intermediate state, the coherent solid-body Lense--Thirring precession of the hot inner flow collapses \citep{2019NewAR..8501524I}. 
Simultaneously, the advancing disk floods the corona with soft seed photons; the resulting coronal cooling deprives the system of the hard Comptonized photons that modulate the oscillation, thereby suppressing and quenching the type-C QPO. A comparable quenching mechanism is discussed by \citet{2027JHEAp..5600738R}; while they note that an inward disk migration towards $r_{\mathrm{ISCO}}$ can theoretically extinguish the precessing flow on a viscous timescale, their observational results attribute the turn-off primarily to coronal cooling and collapse rather than an observed shift in the disk truncation radius. Following the suppression of the horizontally extended precessing flow, the inner accretion geometry reorganizes, transitioning the signal to the type-B QPO observed in epoch 3. In this regime, the horizontally extended hot flow transforms into a more compact, vertically extended structure or base-of-jet configuration \citep{2005ApJ...629..403C, 2016MNRAS.460.2796S}. 
Consequently, the oscillation is no longer driven by global Lense--Thirring precession of the torus, but rather by instabilities or precession near the base of the relativistic jet \citep{2005ApJ...629..403C, 2016MNRAS.460.2796S, 2019NewAR..8501524I}. 
A similar observation has been made for Swift J1727.8–1613, when the type-C QPO and the low-frequency variability got turned off, and a type-B QPO occurred in the dynamic power spectra \citep{2025A&A...699A...9J} at the similar epochs; the source was to be found in the radio flaring state  \citep{2025ApJ...984L..53W}. 

The QPO turning off could be due to the sudden eruption of the jet, which might throw away the materials of the inner coronal region, which was oscillating \citep{2001A&A...370L..17V,2027JHEAp..5600738R}. If the jet has high variability, and it precesses, we can observe the type B QPO. Similarly, for Swift J151857.0$-$572147, when the QPO turned off (\citep{2027JHEAp..5600738R}, a radio activity is reported by \cite{2026arXiv260902681M} using the Merkat data. These observations raise the need for a more detailed understanding of the jet ejection mechanism and its effect on the inner accretion region. A higher-cadence simultaneous monitoring of radio and X-ray can provide stronger evidence of such activity and help understand the origin of type-C and type-B QPOs and their variability properties.


\subsection{Black-hole mass and spin constraints}
\label{sec:mass_spin}

The simultaneous evolution of the low frequency type-C QPO and the
diskbb normalization provides an independent constraint on the BH mass and spin. As the type-C QPO is predominantly associated with 
low-hard state, with the underlying accretion geometry envisaged to having an inner geometrically thick advective flow surrounded by an outer standard Keplerian disk, truncated at 
some radius $R_{\rm o}$. One can then interpret the low-frequency QPO as the global Lense--Thirring (LT) precession of a radially extended,
geometrically thick inner accretion flow with surface-density profile $\Sigma\propto r^{-\zeta}$; the expression as given by \citep{2009MNRAS.397L.101I}

\begin{equation}
\resizebox{\columnwidth}{!}{$
\displaystyle
\nu_{\rm LT} =
\frac{5-2\zeta}{\pi(1+2\zeta)}
\frac{a_*c}{R_g}
\frac{
1-(R_{\rm i}/R_{\rm o})^{1/2+\zeta}
}{
(R_{\rm o}/R_g)^{5/2-\zeta}
(R_{\rm i}/R_g)^{1/2+\zeta}
\left[
1-(R_{\rm i}/R_{\rm o})^{5/2-\zeta}
\right], 
}
$}
\label{eq:lt_physical}
\end{equation}

where, $R_g=\frac{GM_{\rm BH}}{c^2}$.  \, \,  $R_{\rm i}$, here, is the inner radius of the precessing region (i.e., the inner edge of the accretion flow), and $R_{\rm o}$ is the outer radius of the 
precessing region (which is here, the radius of truncation of surrounding outer geometrically thin Keplerian disk). In this scenario, the QPO frequency depends on the
BH mass and spin and on the radial extent of the precessing
flow. The evolution of the diskbb normalization can therefore be
used to estimate the outer radius of the precessing region $(R_{\rm o})$ at each observed QPO frequency. We use two observations of the source for which both the type-C QPO
frequencies and the corresponding diskbb normalizations are available. The two
measurements are
\begin{equation}
\nu_1 = 1.75\pm0.05~{\rm Hz},
\qquad
N_{{\rm diskbb},1}
=11861^{+6305}_{-3402},
\end{equation}
and
\begin{equation}
\nu_2 = 2.50\pm0.05~{\rm Hz},
\qquad
N_{{\rm diskbb},2}
=6716^{+1128}_{-1032}.
\end{equation}

The decrease in the diskbb normalization from the 
first to the second observation corresponds to a decrease 
in the characteristic disk radius ($R_{\rm o}$) . At the same time, the 
QPO frequency increases from 1.75 to 2.50 Hz. This behavior is qualitatively expected if the outer radius 
of the precessing inner flow moves inward during the 
evolution toward the softer state.

\subsubsection{Disk truncation radius from the diskbb normalization}

The physical radius corresponding to the diskbb normalization (which is here the inner disk radius or the truncation radius of the geometrically thin disk) is obtained using the relation from
\begin{equation}
R_{\rm o,j}
=
\xi f_{\rm col}^{2}D_{10}
\left(
\frac{N_{{\rm diskbb},j}}{\cos i}
\right)^{1/2}
{\rm km},
\label{eq:rout}
\end{equation}
where $j=1,2$ denotes the two observations,
$D_{10}=D/(10~{\rm kpc})$, $\xi=0.412$ is the inner-boundary
correction factor\citep{1986ApJ...308..635M,1998PASJ...50..667K}, and $f_{\rm col}=1.7$ is the color-correction
factor \citep{1995ApJ...445..780S}. 
The two radii are therefore not independent free physical
parameters; they are determined by the corresponding diskbb
normalizations, distance, and inclination.

\subsubsection{Mass and spin $-$ dependent inner radius}

The precise location of the inner boundary or edge  of geometrically thick, hot advective accretion flows remains poorly constrained, lacking a well-defined sharp boundary, unlike that of a standard Keplerian disk. Although a few speculative models have been proposed in the literature to define its edge, a definitive physical criteria remains inconclusive [see e.g., 
\citep{2009MNRAS.397L.101I, 2026APJ..1005..238} and references therein]. It is beyond our scope to address this aspect here. As a first approximation, we have considered an accretion flow where the inner radius of the precessing region $(R_{\rm i})$ is assumed to coincide with the last stable circular orbit (i.e., the Kerr-ISCO) [see \citep{2009MNRAS.397L.101I} for more details]. $R_{\rm i}$ will thus contain the explicit information of BH's spin and mass: 

\begin{equation}
\boxed{
R_{\rm i}=R_{\rm ISCO}
=
r_{\rm ISCO}(a_*) \, R_g .
}
\label{eq:isco_physical}
\end{equation}
Putting the values of $G$, $c$, and $M_{\odot}$, one obtains 
\begin{equation}
R_{\rm ISCO}
=
1.476625
\left(\frac{M_{\rm BH}}{M_\odot}\right)
r_{\rm ISCO}(a_*)~{\rm km}.
\end{equation}

The dimensionless ISCO radius depends only on the dimensionless spin
$a_*$ and can be evaluated from the standard expression corresponding to Kerr geometry
\citep{Bardeen1972},
\begin{equation}
r_{\rm ISCO}
=
3+Z_2-
\sqrt{(3-Z_1)(3+Z_2+2Z_1)},
\end{equation}
where, 
\begin{equation}
Z_1 =
1+(1-a_*^2)^{1/3}
\left[
(1+a_*)^{1/3}+(1-a_*)^{1/3}
\right],
\end{equation}
and
\begin{equation}
Z_2=\sqrt{3a_*^2+Z_1^2}.
\end{equation}


\subsubsection{Two LT$-$precession equations}

We make use of the expression given through Eqn. (\ref{eq:lt_physical}), appropriate for our purpose, for which we adopt $\zeta=0.5$ \citep{2009MNRAS.397L.101I}. For the first observation, the measured
diskbb normalization gives $R_{\rm o,1}$ and hence the predicted
frequency can be written as
\begin{equation}
\boxed{
\nu_1^{\rm LT}
=
\nu_{\rm LT}
\left[
M_{\rm BH},a_*,
R_{\rm i}(M_{\rm BH},a_*),
R_{\rm o,1}(D,i,N_1)
\right].
}
\label{eq:lt_state1}
\end{equation}

Similarly, for the second observation, we have 
\begin{equation}
\boxed{
\nu_2^{\rm LT}
=
\nu_{\rm LT}
\left[
M_{\rm BH},a_*,
R_{\rm i}(M_{\rm BH},a_*),
R_{\rm o,2}(D,i,N_2)
\right].
}
\label{eq:lt_state2}
\end{equation}

Using two observed C-type QPO frequencies, these yields 
\begin{equation}
\nu_{\rm LT, 1}(M_{\rm BH},a_*,D,i,N_1)
=
1.75\pm0.05~{\rm Hz},
\label{eq:eq1_massspin}
\end{equation}
and
\begin{equation}
\nu_{\rm LT,2}(M_{\rm BH},a_*,D,i,N_2)
=
2.50\pm0.05~{\rm Hz}.
\label{eq:eq2_massspin}
\end{equation}

Utilizing the diskbb normalizations from Eqns. (3) and (4) associated with 
two type-C QPO observations, one obtains two corresponding different 
$R_{\rm o}$ values. Combining these and adopting an empirically motivated range of values for both the source distance $(D)$ and disk inclination angle $(i)$, one can constrain both the mass and spin of the stellar mass 
BH source using Eqns. (13) and (14). 

\subsubsection{Bayesian treatment of the two equations}

To derive robust constraints on the BH mass $(M_{\rm BH})$ and dimensionless spin $(a_*)$, we employ a Bayesian statistical framework to solve the two frequency constraints simultaneously. The vector of primary physical parameters is defined as: 

\begin{equation}
\boldsymbol{\theta}
=
(M_{\rm BH},a_*,D,i). 
\end{equation}
Meanwhile, the true diskbb normalization of each observation is treated as nuisance parameters, given by $N_1=N_{{\rm diskbb},1}$, and $N_2=N_{{\rm diskbb},2}$. 

We adopt uniform, uninformative, priors spanning broad empirically motivated  ranges to represent our conservative assumptions about the system's geometry and intrinsic properties:

\begin{equation}
3 < \frac{M_{\rm BH}}{M_\odot}<20,
\end{equation}
\begin{equation}
0<a_*<0.998,
\end{equation}
\begin{equation}
3<D<12~{\rm kpc},
\end{equation}
and
\begin{equation}
30^\circ<i<80^\circ.
\end{equation}

For every sampled realization of the parameters, the corresponding physical outer radii $(R_{\rm o})$ are calculated from the sampled diskbb normalizations. The physical ISCO radius is then calculated via Eqn.~(\ref{eq:isco_physical}). To ensure that both observations correspond to physically allowed, radially extended precessing regions, we impose the geometric boundary constraints:

\begin{equation}
R_{\rm o,1}>R_{\rm ISCO} \, \, \, {\rm and} \, \, \, R_{\rm o,2}>R_{\rm ISCO}.
\end{equation}

The joint likelihood of the two QPO) measurements is defined as the product of the individual observation, given by 

\begin{equation}
\mathcal{L}_{\rm joint}
=
\mathcal{L}_1\mathcal{L}_2. 
\end{equation}
Assuming Gaussian measurement uncertainties, each individual likelihood is given by 
\begin{equation}
\mathcal{L}_j
=
\exp\left[
-\frac{1}{2}
\left(
\frac{\nu_{{\rm LT},j}-\nu_{{\rm QPO},j}}
{\sigma_{\nu,j}}
\right)^2
\right].
\end{equation}

Equivalently, the goodness-of-fit can be quantified through the joint \(\chi ^{2}\), computed from the {\rm LT} precession frequencies, given by 
\begin{equation}
\chi^2 =
\left(
\frac{\nu_{{\rm LT},1}-1.75}{0.05}
\right)^2
+
\left(
\frac{\nu_{{\rm LT},2}-2.50}{0.05}
\right)^2.
\label{eq:joint_chi2}
\end{equation}

The asymmetric uncertainties in the two diskbb normalizations are propagated dynamically by treating the measured normalizations as probability distributions rather than fixing them to their best-fit values. The complete joint posterior probability distribution, conditioned on our observational data, is therefore 

\begin{equation}
\begin{aligned}
P(&M_{\rm BH},a_*,D,i\mid{\rm data})
\propto
P(M_{\rm BH})P(a_*)P(D)P(i) \\
&\times
\int dN_1\,dN_2\,
P(N_1|N_{{\rm obs},1})
P(N_2|N_{{\rm obs},2})
\mathcal{L}_1\mathcal{L}_2.
\label{eq:posterior_massspin}
\end{aligned}
\end{equation}

We evaluate this posterior distribution by sampling the adopted prior volume using a scrambled Sobol quasi-Monte Carlo sequence with $2^{21}$ parameter
realizations. The likelihood of every physically allowed realization
is calculated from the two LT-frequency equations. The
likelihood-weighted samples are then marginalized over the source distance,
disk inclination, and the two diskbb normalization nuisance parameters
to obtain the posterior distributions of $M_{\rm BH}$ and $a_*$.

The resulting marginalized posterior gives
\begin{equation}
\boxed{
M_{\rm BH}
=
7.92^{+6.19}_{-3.72}\, M_\odot
}
\end{equation}
and
\begin{equation}
\boxed{a_*=0.163^{+0.116}_{-0.056}. }
\end{equation}

The central values correspond to the median of the posterior
distributions, while the lower and upper uncertainties correspond to
the 16th and 84th percentiles ($1\sigma$  credible intervals ), respectively.

The broad mass uncertainty arises primarily from the degeneracy
between $M_{\rm BH}$, distance, inclination, and the physical
diskbb radius. The spin posterior is more concentrated toward
low positive values because the LT frequency is directly sensitive
to the spin-dependent frame-dragging strength as well as to the
radial extent of the precessing flow.

Herr, the mass and spin are not obtained by fitting the
two observations independently. A single common
$(M_{\rm BH},a_*)$ is required to satisfy both
equations~(\ref{eq:eq1_massspin}) and
(\ref{eq:eq2_massspin}), while the two observations are allowed to
have different outer radii through their measured
diskbb normalizations. The observed increase from
$1.75$ to $2.50$ Hz is therefore incorporated directly into the
joint mass--spin constraint.

The inferred parameters should nevertheless be regarded as
model-dependent estimates. In particular, the original global LT
model allows the inner edge of the hot flow to differ from the ISCO
\citep{2009MNRAS.397L.101I}; here we explicitly impose
$R_{\rm i}=R_{\rm ISCO}$ to connect the precession model with the
Kerr spin. Additional systematic uncertainties associated with the
color-correction factor, inner-boundary correction, distance,
inclination, and the assumed surface-density profile may therefore
affect the derived mass and spin.

It is important to note here that, authors usually prefer Continuum-Fitting (CF) and X-ray Reflection (RF) methods as standard benchmarks for spin estimation. However, the CF method relies heavily on precise independent measurements of mass, distance and inclination angle, while the RF method is sensitive to complex plasma ionization parameters or disk-atmosphere ionization modeling. As most of these parameters are yet unknown to us, we rely on relativistic precession model (RPM) to constrain the spin and mass of the BH source. In fact, the RPM approach utilizes pure general relativistic (GR) timing frequencies, allowing us to solve for mass and spin simultaneously. This kind of method has already been used to estimate the mass and spin of well-known transient sources, such as GRO J1655-40 \citep{2014MNRAS.437.2554M} and XTE J1859+226 \citep{2022MNRAS.517.1469M}. Looking forward, as precise and independent measurements of the system’s distance, mass, and inclination angle become available, future studies can employ the CF and RF methods to independently corroborate our findings. Comparing the spin estimate derived from timing analysis with these independent spectral constraints will provide an important cross-check of strong-field GR and help shed light on the systematic variations observed across the BH spin literature.

\begin{figure*}[!t]
    \centering
    \includegraphics[width=0.6\textwidth]{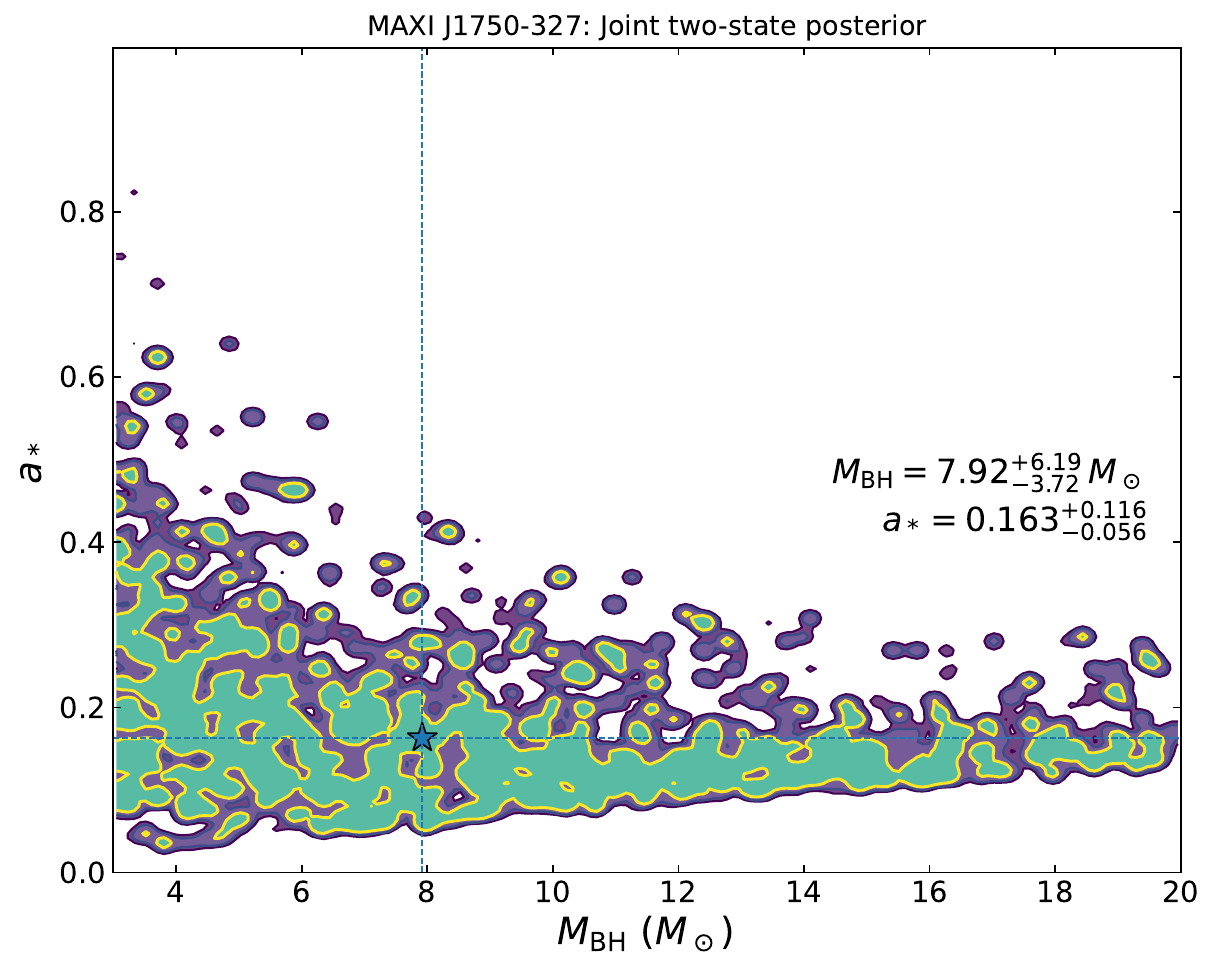}
    \caption{Bayesian posterior probability distribution in the black-hole mass–spin parameter space obtained by jointly fitting the QPO frequencies of the two observed states using the global Lense–Thirring precession model. The filled regions and contour boundaries enclose 68\%, 90\%, and 95\% highest-posterior-density (HPD) credible regions, respectively, calculated from the smoothed two-dimensional posterior density. The black star denotes the posterior median, with the dashed lines marking the corresponding median values of mass and spin. The marginalized constraints give \(M_{\rm BH}=7.92^{+6.19}_{-3.72} \,M_\odot\) and \(a_*=0.163^{+0.116}_{-0.056}\), with the contours indicating the corresponding credible regions.}
\end{figure*}

\section{Conclusions}

\begin{enumerate}

\item A rapid hard-to-soft transition was captured in the first detected outburst of MAXI~J1750-327.
The MAXI/GSC light curve and the HIDs show the $>10$~keV flux declining while the $<10$~keV flux rises over epochs~1--3. A sharp transition occurs between epochs~3 and 4 (MJD~61274--61275), after which the source settles into a compact soft-state region of the HID.

\item The timing evolution is dominated by a drifting QPO that vanishes at the transition. A QPO transition between epochs~2 and 3, likely type-C to type-B, agrees with contemporaneous ATels.

\item The accretion disk moves inward during the transition.
$T_{\rm in}$ rises from 0.72 to a stable 0.92--1.01~keV (epochs~4--11), while $N_{\rm diskbb}$ drops steeply, indicating that the inner disk edge advances toward the ISCO and then stabilizes.

\item The corona cools and is quenched as the source softens.
 $\Gamma$ steepens from 2.38 to 2.93, the non-thermal flux declines, and $F_{\rm nthComp}/F_{\rm diskbb}$ falls from 12 to $<0.5$.

\item Timing--spectral correlations support a Lense--Thirring precession picture.
The QPO frequency correlates positively with $F_{\rm total}$, $\Gamma$, and $T_{\rm in}$, and inversely with $F_{\rm nthComp}/F_{\rm diskbb}$, consistent with $\nu\propto r_{\rm in}^{-3}$ for a shrinking truncation radius. QPO quenching follows once the disk reaches or nears the ISCO and the depleted corona no longer modulates hard photons.

\item The joint Bayesian analysis constrains the black-hole mass to
$M_{\rm BH}=7.92^{+6.19}_{-3.72}\, M_\odot$ and favors a low prograde
spin of $a_*=0.163^{+0.116}_{-0.056}$. The simultaneous increase in
QPO frequency and decrease in diskbb normalization is consistent
with an inward-moving precessing inner flow, although the inferred
mass and spin remain model-dependent.  \\ \\ 

\end{enumerate}

\section*{Acknowledgements}
We gratefully acknowledge the XPoSat project team, facilities team, assembly, integration, and checkout teams, and mission team for their vital contributions to the successful implementation of the XSPECT payload. 
M.H. is grateful to the Department of Physics, University of North Bengal, for providing research facilities and acknowledges the university for awarding a Junior Research Fellowship.
B. R. acknowledges the financial support of the Council of Scientific and Industrial Research (CSIR), India, under the CSIR-SRF Direct fellowship, File No. 09/1022(25279)/2025-EMR-I. S.G. acknowledges IUCAA, Pune, for their support through the Visiting Associateship Program. 

\section*{DATA AVAILABILITY}
This publication uses the data from the \textit{XPoSat} mission of the Indian Space Research Organization (ISRO), archived at the Indian Space Science Data Centre (\href{https://www.issdc.gov.in/}{ISSDC} \url{https://pradan1.issdc.gov.in/x01/}).


\bibliography{sample701}{}
\bibliographystyle{aasjournalv7}



\end{document}